\documentclass[aps,amsmath,amssymb,eqsecnum,preprint,showpacs]{revtex4}
\usepackage{amsfonts,graphicx}

\begin{document}

\title                  {Bianchi II Brane-world Cosmologies (${\cal U}\geq 0$)}

\author                 {R. J.  van den Hoogen}
\email                  {rvandenh@stfx.ca}
\affiliation            {Department of Mathematics, Statistics, and Computer Science,
                        Saint Francis Xavier University, Antigonish, N.S.,
                        B2G 2W5, Canada}

\author                 {J. Iba{\~n}ez}
\email                  {wtpibmej@lg.ehu.es}
\affiliation            {Departamento de F{\'i}sica Te{\'o}rica,
                        Universidad del Pa{\'i}s Vasco, Bilbao, Spain}

\begin{abstract}
The asymptotic properties of the Bianchi type II cosmological model in the Brane-world scenario are investigated.  The matter content is assumed to be a combination of a perfect fluid and a minimimally coupled scalar field that is restricted to the Brane.  The isotropic braneworld solution is determined to represent the initial singularity in all brane-world cosmologies.  Additionally, it is shown that it is the kinetic energy of the scalar field which dominates the initial dynamics in these brane-world cosmologies. It is important to note that, the dynamics of these brane-world cosmologies is not necessarily asymptotic to general relativistic cosmologies to the future in the case of a zero four-dimensional cosmological constant.

\end{abstract}

\pacs{PACS numbers(s): 04.20.Jb, 98.80.Hw}

\maketitle{}


\section{Introduction}
 
It is theorized that Einstein's General Relativity breaks down at sufficiently high energies. Therefore, General Relativity may not be the most appropriate choice for a gravitational theory in the very early universe.   
Developments in string theory suggest that gravity may truly be a higher dimensional theory, becoming an effective 4-dimensional theory at lower energies.  Some researchers have suggested an alternative scenario in which the matter fields are restricted to a $3$-dimensional brane-world embedded in $1+3+d$ dimensions (the bulk), while the gravitational field is free to propagate in the $d$ extra dimensions \cite{Rubakov83,Antoniadis98,Binetruy00a}.  This new scenario for the gravitational field will have significant implications for cosmology.

The dynamical equations on the 4-dimensional brane-world differ from the field equations of General Relativity, since there exist additional terms that carry non-local gravitational effects from the bulk onto the brane.  These additional terms are quadratic in nature and reduce to the regular Einstein Field equations of General Relativity for late times.  However, for early times, these quadratic terms will play a very critical role in the evolutionary dynamics of these brane-world models.

 A lot of effort has been directed at the so-called perfect fluid Friedmann Brane-world models \cite{Binetruy00a,Kaloper99,Mukohyama00,Binetruy00b,Campos01a,Campos01b}.  More recently, people have begun to investigate anisotropic brane-world models \cite{Campos01a,Campos01b,Maartens01,Toporensky01,Coley02,Russians}, trying to determine the effects of the bulk gravitational field.  People have also begun to look at the effect of scalar fields in the Friedmann brane-world \cite{Goheer02}.  It is known in General Relativity that only sets of measure zero in the set of all spatially homogeneous models (satisfying a set of reasonable energy conditions) will isotropize \cite{CollinsHawking73}.  It is not yet clear whether this result holds for spatially homogeneous brane-world models as well.

General relativistic cosmological models containing both a perfect fluid and a scalar field having an exponential potential have been previously studied \cite{Billyard99,vandenHoogen99,Billyard98}.  One of the exact solutions found in these models is   a spatially flat isotropic  model with the property that the energy density of the scalar field is proportional to the energy density of the perfect fluid, and consequently are labeled scaling cosmologies (See \cite{Wands97,Copeland98} and references therein).  It was found in \cite{Billyard99,vandenHoogen99,Billyard98} that these cosmological scaling solutions are in general not stable, and hence do not represent typical late-time behaviour.  However, during the evolution of one of these cosmological models containing both a perfect fluid and a scalar field having an exponential potential, these scaling solutions may play an important role in the intermediate or transient behaviour of these cosmologies, since the scalar field and the perfect fluid are non-negligible in a neighborhood of the scaling solution.  In particular,  in these models a significant portion of the current energy density of the Universe could be contained in the scalar field whose dynamical effects mimic those of cold dark matter.

In general relativistic cosmological models, the Bianchi II model plays an important role in the past asymptotic behaviour of the Bianchi IX models.  Recently, Coley \cite{Coley02} has argued that the Bianchi type IX model has an isotropic singularity in the brane-world scenario.  This indicates that the Bianchi II model in the brane-world scenario may also have a behaviour that is different than in the general relativistic case.  In this paper, we shall assume that the brane-world is a spatially homogeneous Bianchi type II model containing both a scalar field and a perfect fluid.  Consequently, due to the additional terms that carry non-local gravitational effects from the bulk onto the brane, new evolutionary behaviours in the cosmologies are expected. In particular, with a combination of a perfect fluid and a scalar field, due to the quadratic nature of the non-local gravitational effects, a wide variety of new and exotic asymptotic and transient behaviours are possible.

Maartens \cite{Maartens00} and Shiromizo et al. \cite{Shiromizu00,Sasaki00} have developed an elegant covariant approach to the bulk effects on the brane.  The equations derived by Maartens are an extension of earlier work by Ellis and MacCallum \cite{EllisMacCallum69} that has been subsequently developed more recently in the book by Wainwright and Ellis \cite{EllisWainwright97}.   Using the formalism developed by Maartens and Wainwright and Ellis, we propose to investigate the dynamical behavior in a wider class of anisotropic models than what has been previously analyzed.

The resulting field equations will yield a system of ordinary differential equations, suitable for a geometric analysis using dynamical systems techniques.   This analysis will determine whether the dynamics of the brane-world scenario mimics the dynamics of a General Relativistic cosmology or at the very least, is asymptotic to a general relativistic cosmology.  One would be interested in both the early (nature of the initial singularity) and late time behavior (i.e., whether these models approach the General Relativistic regime and isotropize).

With regards to notation, quantities having a tilde are objects found in the five dimensional bulk while non-tilde quantities are objects restricted to the four dimensional brane-world.  The upper case latin subscipts $(A,B,C,\dots)$ range from $0..4$, the lower case latin subscripts $a,b,c,\dots)$ range from $0..3$ and the greek subscripts $(\alpha,\beta, \gamma,\dots)$ range from $1..3$.  


\section{Governing Equations}

The popular brane-world scenario assumes the existence of a domain wall, $M$, (a 3-brane)  with induced metric $g_{AB}=\tilde g_{AB}-\tilde n_A \tilde n_B$ in a five dimensional spacetime $\widetilde M$ with metric $\tilde g_{AB}$ where $\tilde n^A$ is the spacelike unit vector normal to $M$. In a neighborhood of the brane, one can choose coordinates $x^{A}=(x^a,\chi)$ such that $\tilde n_A=\chi_{,A}$ where $\chi=0$ coincides with the brane. The $x^a$ are the spacetime coordinates on the brane $M$. The five dimensional field equations are Einstein's equations with a bulk cosmological constant $\widetilde\Lambda$, and a brane energy-momentum source $T_{AB}$, 
$$\tilde G_{AB}=\widetilde\kappa^2[-\widetilde\Lambda g_{AB} +\delta(\chi)\{\lambda g_{AB}+ T_{AB}\}]$$
where $\widetilde\kappa^2=8\pi/\widetilde M_{\rm p}^{\,3}$, with $\widetilde M_{\rm p}$ being the five-dimensional Plank mass and $\lambda$ is the brane tension.  

The field equations induced on the brane have been derived using a geometric approach by Shiromizu et al. \cite{Shiromizu00,Sasaki00,Maartens00}, using
the Gauss-Codazzi equations, Israel's junction condition and imposing a $Z_2$
symmetry with the brane as a fixed point. The result is a modification of the standard Einstein equations with new terms ($S_{ab}$ and  $ {\cal E}_{ab}$ which are defined later) carrying bulk effects onto the
brane:
\begin{equation}
G_{ab}=-\Lambda g_{ab}+\kappa^2
T_{ab}+\widetilde{\kappa}^4S_{ab} - {\cal E}_{ab}
\end{equation}
where 
\begin{equation}
\kappa^2=\frac{8\pi}{M_{\rm p}^2}\qquad
\lambda=6{\frac{\kappa^2}{\widetilde\kappa^4}}\qquad
\Lambda=\frac{4\pi}{ \widetilde{M}_{\rm p}^{\,3}}\left[\widetilde{\Lambda}+
\left({\frac{4\pi}{3\widetilde{M}_{\rm p}^{\,3}}}\right)\lambda^2\right]
\end{equation}
with $M_{\rm p}$ being the four-dimensional Plank mass, and $\Lambda$ is the effective four-dimensional cosmological constant on the Brane. It is common to assume that the effective cosmological constant, $\Lambda$, on the brane is zero through fine tuning of the brane tension $\lambda$. However,
we shall assume for the moment that it is non-zero.

We will assume that the source term, $T_{ab}$ (restricted to the Brane) is a non-interacting mixture of ordinary matter having energy density $\rho$, and a minimally coupled scalar field, $\phi$, that is  
\begin{equation}
T_{ab}=T_{ab}^{\ \ perfect \ fluid}+T_{ab}^{\ \ scalar\ field}.\label{source}
\end{equation}
We will assume that  the matter component is equivalent to a co-moving perfect fluid with 4-velocity $u^a$ and a linear barotropic equation of state $p = (\gamma-1)\rho$. Energy conditions  impose the
restriction $\rho\geq 0$ and causality requires that $\gamma\in[0,2]$. We will assume that the potential for the scalar field has an exponential form, that is, $V(\phi)=V_0e^{k\kappa\phi}=V_0e^{\frac{\sqrt{8\pi}}{M_p}k\phi}$. 
The energy-momentum tensors restricted to the brane for a perfect fluid and a minimally-coupled scalar field are  
\begin{eqnarray}
T_{ab}^{\ \ perfect \ fluid}&=&\rho u_a u_b+ph_{ab},\\
T_{ab}^{\ \ scalar\ field} &=& \phi_{;\,a}\phi_{;\,b}
-g_{ab}\left(\frac{1}{2}\phi_{;\,c}\phi^{;\,c}+V(\phi)\right),
\end{eqnarray}
where the projection tensor $h_{ab} \equiv g_{ab}+u_a u_b$ projects orthogonal to $u^a$.   We note that if $\phi^{;\,a}$ is timelike, then a scalar field with potential $V(\phi)$ is equivalent to a perfect fluid having an energy density and isotropic pressure 
\begin{eqnarray}
\rho^{scalar\ field}&=&-\frac{1}{2}\phi_{;\,c}\phi^{;\,c}+V(\phi),\\
p^{scalar\ field} &=&-\frac{1}{2}\phi_{;\,c}\phi^{;\,c}-V(\phi).
\end{eqnarray}

The bulk corrections to the Einstein equations on the brane are of
two forms: firstly, the matter fields contribute local quadratic
energy-momentum corrections via the tensor $S_{ab}$, and
secondly, there are nonlocal effects from the free gravitational
field in the bulk. The local matter corrections are
given by
\begin{equation}
S_{ab}=\frac{1}{12}T_c{}^c T_{ab}
-\frac{1}{4}T_{ac}T^c{}_b+ \frac{1}{24}g_{ab} \left[3 T_{cd}
T^{cd}-\left(T_c{}^c\right)^2 \right].
\end{equation}
Since the $S_{ab}$ is essentially quadratic in the energy momentum tensor, we should expect cross terms between the scalar field and the perfect fluid.   In this paper we shall assume that the gradient of the scalar field $\phi^{;\,a}$, is aligned with the fluid 4-velocity, $u^{a}$, that is $\phi^{;\,a}/\sqrt{-\phi_{;\,b}\phi^{;\,b}}=u^a$. In general $\phi^{;\,a}$ need not be aligned with $u^a$ thereby creating a rich variety of cross terms. The local brane effects due to a combination of a perfect fluid and a scalar field are  
\begin{eqnarray}
S_{ab} &=&\frac{1}{12} \left(\rho-\frac{1}{2}\phi_{;\,c}\phi^{;\,c}+V(\phi)\right)^2 u_a u_b\nonumber\\
&&+\frac{1}{12}\left(\rho-\frac{1}{2}\phi_{;\,c}\phi^{;\,c}+V\right)
\left(\rho+2 p-\frac{3}{2}\phi_{;\,c}\phi^{;\,c}-V(\phi)\right)h_{ab}
\end{eqnarray}
 
The non-local effects from the free gravitational field in the bulk are characterized by the projection of the bulk Weyl tensor onto the brane.  Given a timelike congruence on the brane, the bulk correction, ${\cal
E}_{ab}$ can be decomposed \cite{Maartens00} via
\begin{equation}
{\cal E}_{ab}=-\left(\frac{\widetilde \kappa}{\kappa}\right)^4\left[{\cal U}(u_{a}u_{b}+\frac{1}{3}h_{ab})+{\cal P}_{ab}+ 2{\cal Q}_{(a}u_{b)}\right]
\end{equation}
(See \cite{Maartens00} for further details.)  In general, the conservation equations (the contracted Bianchi identities on the brane) do not determine all of the independent components of ${\cal E}_{ab}$. In particular, there is no
equation to determine ${\cal P}_{ab}$ and hence, in general, the projection of the
5-dimensional field equations onto the brane do not lead to a
closed system of equations.  In the cosmological context,  in which the background metric is
spatially homogeneous and isotropic, we have that
\begin{equation}\label{bulk_weyl}
{\rm D}_a{\cal U}={\cal Q}_{a}={\cal P}_{ab}=0
\end{equation}
where ${\rm D}_a$ is the totally projected part of the brane
covariant derivative.
Since ${\cal P}_{ab}=0$, in this case the evolution of ${\cal E}_{ab}$ is fully determined \cite{RandallSundrum99,Arkani-Hamed00,ChamblinGibbons00}. 

All of the brane source terms and bulk corrections mentioned above may be consolidated into an effective total energy density, pressure, energy flux, and anisotropic pressure as follows. The modified Einstein equations take the standard Einstein form with a redefined energy-momentum tensor:
\begin{equation}
G_{ab}=\kappa^2 T^{\ \ \rm total}_{ab}
\end{equation}
where
\begin{equation}
T^{\ \ \rm total}_{ab} \equiv-\frac{\Lambda}{\kappa^2}g_{ab}+ T_{ab}+\frac{\widetilde{\kappa}^{4}}{\kappa^2}S_{ab}
- \frac{1}{\kappa^2}{\cal E}_{ab}. 
\end{equation}
The total equivalent energy density, pressure, energy flux, and anisotropic pressure due to a perfect fluid,  scalar field, and both local and non-local brane effects are
\begin{eqnarray} 
\rho^{\rm total} \label{total1} &=&\frac{\Lambda}{\kappa^2}+ \rho+\left(-\frac{1}{2}\phi_{;\,a}\phi^{;\,a}+V(\phi)\right)
 \nonumber\\&& \qquad+\frac{\widetilde{\kappa}^{4}}{\kappa^6}
\Biggl[\frac{\kappa^4}{12}\left(\rho-\frac{1}{2}\phi_{;\,a}\phi^{;\,a}+V(\phi)\right)^2+{\cal U}\Biggr] \\
p^{\rm total} \label{total2} &=& -\frac{\Lambda}{\kappa^2}+p+\left(-\frac{1}{2}\phi_{;\,a}\phi^{;\,a}-V(\phi)\right)\nonumber\\
&&+\frac{\widetilde{\kappa}^{4}}{\kappa^6}
\Biggl[ \frac{\kappa^4}{12}\left(\rho-\frac{1}{2}\phi_{;\,a}\phi^{;\,a}+V(\phi)\right)
\left(\rho+2p-\frac{3}{2}\phi_{;\,a}\phi^{;\,a}-V(\phi)\right)
+\frac{1}{3}{\cal U}\Biggr]\\
q_{a}^{\rm total} \label{total3} &=& \frac{\widetilde \kappa^4}{\kappa^6}{\cal Q}_{a}\\
\pi_{ab}^{\rm total} \label{total4} &=& \frac{\widetilde \kappa^4}{\kappa^6}{\cal P}_{ab} 
\end{eqnarray}

As a consequence of the form of the bulk energy-momentum tensor
and the $Z_2$ symmetry, it follows \cite{Shiromizu00,Sasaki00} that
the brane energy-momentum tensor separately satisfies the
conservation equations, (where we have tacitly assumed that the scalar field and the matter are non-interacting) that is,
\begin{eqnarray}
T^{a\ \ perfect\ fluid}_{b;\,a}&=&0  \label{matter_conservation},\\
T^{a\ \ scalar\ field}_{b;\,a}&=&0  \label{Klein-Gordon}.
\end{eqnarray}
Given that $T^{a}_{b;\,a}=0$, the contracted Bianchi identities on the brane imply that the projected Weyl tensor obeys the constraint
\begin{equation}
{\cal E}^{a}_{b;\,a}=\widetilde{\kappa}^4 S^{a}_{b;\,a}.
\label{E_dot}
\end{equation}


\section{Evolution and Constraint Equations for the Bianchi II Brane-world} 

\subsection{Orthonormal Frame Formalism}

We shall adopt the orthonormal frame formalism as developed by Ellis and MacCallum \cite{EllisMacCallum69} and further developed in \cite{EllisWainwright97}.  We shall assume that there is a $G_3$ group of motions  acting simply transitively on three dimensional spacelike hypersurfaces.   We will choose an orthonormal frame of vector fields 
$\{{\bf e}_a\}$ and align ${\bf e}_0$ to the fluid 4-velocity ${\bf u}$, i.e., ${\bf e}_0={\bf u} $. Essentially we are assuming that the velocity field of the matter source is non-tilted with respect to the normal vector field of the hypersurfaces generated by the $G_3$ group of motions.  Therefore, the triad of spacelike vectors $\{{\bf e}_\alpha\}$ at each point spans the tangent space  of the group orbits and hence coordinates can be chosen such that the commutation functions are functions of $t$ only.

The commutation relation between ${\bf e}_0$ and ${\bf e}_\alpha$ yields the kinematic quantities associated with the vector field $\bf u$. That is
$$
[{\bf e}_0,{\bf e}_\alpha]=\gamma^{b}_{0 \alpha}{\bf e}_b
$$
where 
$$\gamma^{0}_{0\alpha}=\dot u_{\alpha},\qquad{\rm and} \qquad \gamma^{\beta}_{0\alpha}=-\sigma_\alpha^{\ \beta}-H\delta_\alpha^{\ \beta}-\epsilon_{\alpha\delta}^{\ \ \beta}(\omega^\delta+\Omega^\delta).$$
Where $H$ can be interpreted as the expansion of the fluid, $\sigma_{\alpha\beta}$ can be interpreted as the rate of shear tensor of the fluid, $\dot u_{\alpha}$ can be interpreted as the acceleration of the fluid, and $\omega_\alpha$ is a measure of the vorticity of the fluid. The quantity $\Omega_\alpha$ can be interpreted as the angular velocity of the spatial tetrad $\{{\bf e}_\alpha\}$ with respect to a Fermi-propagated spatial frame.  However, with the assumptions made thus far both $\dot u_{\alpha}=0$ and $\omega_\alpha=0$.

The vector fields  ${\bf e}_\alpha$ generate a Lie Algebra with commutation functions $\gamma^{\delta}_{\alpha\beta}$ that can be decomposed in the following way
\begin{eqnarray}
[{\bf e}_\alpha,{\bf e}_\beta] &=& \gamma^\delta_{\alpha\beta}{\bf e}_\delta\\
 &=& \biggr(\epsilon_{\alpha\beta\gamma}n^{\delta\gamma}
+a_\alpha\delta^\delta_\beta-a_\beta\delta^\delta_\alpha\biggl){\bf e}_\delta
\end{eqnarray}
where $n_{\alpha\beta}$ and $a_{\alpha}$ are functions of time $t$ only.  One can use a time dependent spatial rotation to diagonalize $n_{\alpha\beta}={\rm diag}(n_1,n_2,n_3)$ \cite{EllisMacCallum69,EllisWainwright97}.  The Jacobi identity $[{\bf e}_1,[{\bf e}_2,{\bf e}_3]]=0$ (and permutations) imply that $n_{\alpha\beta}a^\beta=0$, motivating the classification of a $G_3$ group of motions into two classes, Class A where $a^\alpha=0$, and Class B where $a^\alpha \not =0$.  These two classes be be further sub-classified  to yield the nine Bianchi types.  

In this paper we are interested in a set of Class A models and in particular the Bianchi type II models.  For the Bianchi type II models only one of the $n_{\alpha\beta}$ is different from zero.  Here we assume that $n_1=n$, and $n_2=n_3=0$.  In addition, we shall assume that the spatial triad $\{{\bf e}_\alpha\}$ is Fermi-propagated which implies that $\Omega_\alpha=0$.  In choosing a Fermi-propagated triad, we have effectively determined the effective energy flux, that is, $q^{\rm total}_\alpha=0$. This then imposes the condition that ${\cal Q}_\alpha=0$.  In addition, in choosing a Fermi-propogated frame, one now has the freedom to rotate the spatial frame so that $\sigma_{\alpha\beta}={\rm diag}(\sigma_{11},\sigma_{22},\sigma_{33})$ \cite{EllisMacCallum69}.  Since there is no way to determine ${\cal P}_{ab}$ from observations on the brane, we impose the condition that our brane world model should be consistent with a Friedmann-LeMaitre cosmological brane model by assuming that ${\cal P}_{\alpha\beta}=0$. The equations describing the evolution of this class of  models can be found in equations 1.90--1.100 of \cite{EllisWainwright97}.  The Einstein Field equations and Jacobi identities yield
\begin{eqnarray}\label{field-equations}
\dot H &=& -H^2-\frac{2}{3}\sigma^2-\frac{\kappa^2}{6}(\rho^{\rm total}+3p^{\rm total})\\
\dot \sigma_{11} &=& -3H\sigma_{11}-\frac{2}{3}n^2\\
\dot \sigma_{22} &=& -3H\sigma_{22}+\frac{1}{3}n^2\\
\dot \sigma_{33} &=& -3H\sigma_{33}+\frac{1}{3}n^2\\
\dot n &=& (-H+2\sigma_{11})n\\
\kappa^2\rho^{\rm total} &=& 3H^2-\sigma^2-\frac{1}{4}n^2
\end{eqnarray}
where $\rho^{\rm total}$, and $p^{\rm total}$ are defined in equations \ref{total1}--\ref{total2} and we have introduced the constant $\kappa^2$. Note, not all of the above equations are independent in particular $\sigma_{11}+\sigma_{22}+\sigma_{33}=0$.

From the conservation equations on the brane, (see equations \ref{matter_conservation} and \ref{Klein-Gordon}) we obtain
\begin{eqnarray}
\dot \rho & = & -3\gamma H \rho \\
\ddot \phi &=& -3H\dot\phi - k\kappa V(\phi)
\end{eqnarray}
Furthermore from the Bianchi identities on the brane (see equations \ref{E_dot} and \ref{bulk_weyl}) we obtain (see \cite{Maartens00}),
\begin{equation}
\dot {\cal U} = -4H{\cal U}.\label{U-dot}
\end{equation}

\subsection{Dimensionless Formalism}

To facilitate the analysis of these models and to compare with previous work we now choose new variables of the form

\begin{equation}\label{new-variables}
\begin{tabular}{lll}
$\displaystyle \Sigma_+         =\frac{1}{2}\frac{\sigma_{22}+\sigma_{33}}{H}$,          \qquad\qquad\qquad &
$\displaystyle\Sigma_-          =\frac{1}{2\sqrt{3}}\frac{\sigma_{22}-\sigma_{33}}{H}$,  \qquad\qquad\qquad &
$\displaystyle N                =\frac{1}{2\sqrt{3}}\frac{n}{H}$,                  \\
\\
$\displaystyle \Omega_{pf}      =\frac{\kappa^2\rho}{3H^2}$,                            &
$\displaystyle \Omega_{\Lambda} = \frac{\Lambda}{3H^2}$,                          &
$\displaystyle \Omega_{\cal U}  =\frac{\widetilde \kappa^4}{\kappa^4}\frac{\cal U}{3H^2}$,\\
\\
$\displaystyle \Phi             =\frac{\kappa^2V}{3H^2}$,                         &
$\displaystyle \Psi             =\sqrt{\frac{3}{2}}\frac{\kappa\dot\phi}{3H}$,       \\
\\
\multicolumn{3}{l}{$\displaystyle \Omega_{\lambda}   
        =\frac{\widetilde\kappa^4}{36H^2}\left(\rho+\frac{1}{2}\dot\phi^2+V\right)^2 
        =\frac{\kappa^2}{6\lambda H^2}   \left(\rho+\frac{1}{2}\dot\phi^2+V\right)^2 =\frac{3H^2}{2\kappa^2\lambda}\left(\Omega_{pf}+\Psi^2+\Phi\right)^2 $} 
\end{tabular}
\end{equation}
and a new time variable $$\frac{dt}{d\tau}=\frac{1}{H}.$$

The system of equations (\ref{field-equations}--\ref{U-dot}) becomes (where prime denotes differentiation with respect to $\tau$).
\begin{eqnarray}\label{DS}
\Sigma_+'           &=& \Sigma_+(q-2)+4N^2 \\
\Sigma_-'           &=& \Sigma_-(q-2) \\
N'                  &=& N(q-4\Sigma_+)\\
\Omega_{pf}'        &=& \Omega_{pf}\left(2q-(3\gamma-2)\right)\\
\Omega_{\Lambda}'   &=& 2\Omega_{\Lambda}(q+1)\\
\Omega_{\lambda}'    &=& \Omega_{\lambda}\left(2q-(3\gamma_{\lambda}-2)\right)\label{lambda-prime}\\
\Omega_{\cal U}'    &=& 2\Omega_{\cal U}(q-1)\label{U-prime}\\
\Psi'               &=& \Psi(q-2)-\frac{\sqrt{6}}{2}k\Phi \\
\Phi'               &=& 2\Phi(q+1+\frac{\sqrt{6}}{2}k\Psi)
\end{eqnarray}
where 
\begin{eqnarray}
q &\equiv& 2\Sigma_+^2+2\Sigma_-^2
+\frac{3\gamma-2}{2}\Omega_{pf}-\Omega_{\Lambda}+\frac{3\gamma_{\lambda}-2}{2}\Omega_{\lambda}+\Omega_{\cal U}+2\Psi^2-\Phi\, ,
\nonumber\\
\gamma_{\lambda} &\equiv& \frac{2\gamma\Omega_{pf}+4\Psi^2}{\Omega_{pf}+\Psi^2+\Phi}\,, 
\end{eqnarray}
where the generalized Friedmann equation becomes
\begin{equation}
\Omega^{\rm total}\equiv\Omega_{pf}+\Omega_{\Lambda}+\Omega_{\lambda}+\Omega_{\cal U}+\Psi^2+\Phi= 1-\Sigma_+^2-\Sigma_-^2-N^2.\label{constraint}
\end{equation}
 In addition, the equation for $H$ decouples  and becomes 
 \begin{equation}
 H'=-(1+q)H
 \end{equation}
 
We have now determined the equations describing the evolution of the Bianchi II brane-world model.  The resulting equations are suitable for a qualitative analysis using techniques from dynamical systems theory. In general, if we let ${\bf X}=[\Sigma_+,\Sigma_-,N,\Omega_{pf}, \Omega_{\Lambda},\Omega_{\lambda},\Omega_{\cal U},\Psi,\Phi]$, then the system of equations (\ref{DS}), can be interpreted as ${\bf X}'={\bf F}({\bf X})$ where ${\bf F}:{\bf X}\subset \mathbb{R}^9 \to \mathbb{R}^9$. Please see review of dynamical systems theory in \cite{EllisWainwright97} and other texts such as \cite{Perko91,Wiggins90} for a more complete description of dynamical systems techniques applied to systems of ordinary differential equations.   


\section{ Qualitative Analysis of the Bianchi II Brane World Models}

\subsection{Invariant Sets}

We note that equation (\ref{constraint}) is a constraint or conservation equation that we rewrite as
\begin{equation}
G({\bf X})\equiv  1-\Sigma_+^2-\Sigma_-^2-N^2 -\Omega_{pf} - \Omega_{\Lambda}- \Omega_{\cal U}-\Omega_{\lambda}-\Phi-\Psi^2=0.\label{constraint2}
\end{equation}  
Hence the phase space of the dynamical system $\cal S$ is an eight dimensional subset of ${\mathbb R}^9$, where  
$${\cal S} = \{ {\bf X}\in {\mathbb R}^9|G({\bf X})=0\}.$$
Here, one is able to use (\ref{constraint2}) to eliminate the variable $\Omega_{\Lambda}$ globally from the dynamical system (\ref{DS}), which results in a system of eight differential equations.  

The dynamical system (\ref{DS}) is invariant under the transformation $N \to -N$, therefore without loss of generality we restrict the analysis to the set $N\geq 0$.  The set $N^+=\{{\bf X}\in {\cal S}| N>0\}$ represents Bianchi II models, where as, the set $N^0=\{{\bf X}\in {\cal S}| N=0\}$ represents the Bianchi I models (if $\Sigma_+^2+\Sigma_-^2\not = 0$) or the zero curvature Friedman-Lemaitre models (if $\Sigma_+^2+\Sigma_-^2= 0$).

The dynamical system (\ref{DS}) is also invariant under the transformation $\Sigma_- \to -\Sigma_-$, therefore without loss of generality we restrict the analysis to the set $\Sigma_-\geq 0$. 

The evolution equation for $\Omega_{\cal U}$ implies that the surface $\Omega_{\cal U}=0$ divides the phase space into three distinct regions, ${\cal U}^+=\{{\bf X}\in {\cal S}|\Omega_{\cal U}>0\}$, ${\cal U}^0=\{{\bf X}\in {\cal S}| \Omega_{\cal U}=0\}$ and ${\cal U}^-=\{{\bf X}\in {\cal S}| \Omega_{\cal U}<0\}$.  

With the assumption of the weak energy condition (i.e., $\rho_{pf}+\frac{1}{2}\dot\phi^2 +V \geq 0$), it can be shown that in the invariant sets ${\cal U}^+$ and ${\cal U}^0$ that 
$$0 \leq \Sigma_+^2,\Sigma_-,N, \Omega_{pf},\Omega_{\Lambda},\Omega_{\lambda},\Omega_{\cal U},\Psi^2,\Phi,\leq 1,$$
that is, ${\cal U}^+$ and ${\cal U}^0$ are compact subsets of ${\cal S}$.
Unfortunately, the invariant set ${\cal U}^-$ is not compact in the variables we have chosen.  One can however choose different dimensionless variables (\ref{new-variables}) to remedy this problem (see van den Hoogen and Abolghasem \cite{vandenHoogen02b}).

There are various invariant sets associated with the matter content.  We define six sets
\begin{eqnarray*}
^0\Omega^{0,0}&=&\{{\bf X}\in {\cal S}| \Omega_{pf}=0, \Phi=0, \Psi=0 \},\\
^0\Omega^{0,\pm}&=&\{{\bf X}\in {\cal S}| \Omega_{pf}=0, \Phi=0, \Psi\not =0 \},\\
^0\Omega^{+,\pm}&=&\{{\bf X}\in {\cal S}| \Omega_{pf}=0, \Phi\not =0, \Psi\not = 0 \},\\
^+\Omega^{0,0}&=&\{{\bf X}\in {\cal S}| \Omega_{pf}\not=0, \Phi=0, \Psi=0 \},\\
^+\Omega^{0,\pm}&=&\{{\bf X}\in {\cal S}| \Omega_{pf}\not=0, \Phi=0, \Psi\not =0 \},\\
^+\Omega^{+,\pm}&=&\{{\bf X}\in {\cal S}| \Omega_{pf}\not=0, \Phi\not =0, \Psi\not = 0 \},
\end{eqnarray*}
where the notation is interpreted as  $$ ^{({\rm value\ of\ } \Omega_{pf})} \Omega ^{({\rm value\ of\ }\Phi,{\rm value\ of\ }\Psi)}$$

One additional invariant set of interest is the invariant set ${\cal GR}=\{{\bf X}\in {\cal S}| \Omega_{\lambda} = 0,\Omega_{\cal U} = 0\}$.  One is interested in whether this set is the future attractor of all brane-world cosmologies.

\subsection{Monotonic Functions}\label{monotonic}

In the set ${\cal U}^+$, $\Omega_{\Lambda}>0$, it can be shown that $q>-1$.  Therefore we easily see that $\Omega_{\Lambda}$ monotonically increases from its lower limit $\Omega_{\Lambda} = 0$ to its upper limit $\Omega_{\Lambda}=1$.  Furthermore, with the existence of this monotonic function, we can rule out the existence of any closed or periodic orbits in the full eight-dimensional phase space, ${\cal S}$.  That is, any closed or periodic orbits (if they exist) must lie on the lower-dimensional boundaries of ${\cal S}$.

The seven-dimensional set ${\cal U}^0$, $\Omega_{\Lambda}>0$ lies on the boundary of ${\cal S}$.  If $\Sigma_- \not = 0$ (or $\Omega_{pf}\not = 0$)  it can be shown that $q>-1$ and again $\Omega_{\Lambda}$ monotonically increases.   Again the existence of this monotonic function implies that there do not exist any closed or periodic orbits in this set.  Note the only invariant subset of ${\cal U}^0$, $\Omega_{\Lambda}>0$ in which $q=-1$ (i.e, the only invariant subset of ${\cal U}^0$ in which  $\Omega_{\Lambda}'=0$) is $\Sigma_+=\Sigma_-=N=\Omega_{pf}=\Psi=\Phi=0$ and $\Omega_{\Lambda}+\Omega_{\lambda}=1$.

With the arguments made in the previous two paragraphs, the only sets in which  a periodic and/or closed orbit might exist is the seven-dimensional set $\Omega_{\Lambda}=0$ and the five-dimensional set $\Omega_{pf}=\Omega_{\cal U}=\Sigma_-=0,\Omega_{\Lambda}>0$.  In all other cases we observe that the future asymptotic behaviour characterized by $\lim_{\tau \to \infty}\Omega_{\Lambda} = 1$ (this also implies through the constraint equation (\ref{constraint}) that $\lim_{\tau \to \infty} {\bf X} = [0,0,0,0,1,0,0,0,0]$), the DeSitter Model (see local analysis in next section), while the past asymptotic behaviour is characterized by $\lim_{\tau \to -\infty}\Omega_{\Lambda} = 0$, cosmological models with a four dimensional cosmological constant that is equivalently equal to zero.

\subsection{Equilibrium Points and Local Behaviour} 

The equilibrium points can be classified in to one of the six invariant sets $^{(\ )}\Omega^{(\ ,\ )}$, depending on the matter content.  The order of the coordinates is $[\Sigma_+,\Sigma_-,N,\Omega_{pf},\Omega_{\Lambda},\Omega_{\lambda},\Omega_{\cal U},\Psi,\Phi])$.  In what follows, for those points located in the $\Omega_{\Lambda}=0$ invariant set, the eigenvalue found in brackets $\{\ \}$ is the eigenvalue associated with $\Omega_{\Lambda}$ direction.

\subsubsection{Vacuum, $^0\Omega^{0,0}$}\label{vacuum}

Unfortunately the dynamical system (\ref{DS}) is not differentiable everywhere, and in particular, it is not differentiable in a neighborhood of this invariant set.  However, since the system is continuous, we are able to find the equilibrium points of the dynamical system.  To determine the stability of each equilibrium point in this set, we change the variables $(\Omega_{pf},\Psi,\Phi)$ to spherical coordinates $(\tilde\rho, \tilde\theta, \tilde\phi)$ and analyze the resulting system.  This is done in Appendix \ref{Appendix_V}.

\begin{description}
 
\item {$DS$;\ De-Sitter}  $[0,0,0,0,1,0,0,0,0]$.  

\item{$m$;\ Isotropic Braneworld Solution} $[0,0,0,0,0,1,0,0,0]$. 

\item{$R^0$;\ Robertson Walker Dark Radiation Solution }   
$[0,0,0,0,0,0,1,0,0]$.   

\item{$R^{II}$;\ Bianchi II Dark Radiation Solution }   
$[\frac{1}{4},0,\frac{1}{4},0,0,0,\frac{7}{8},0,0]$.   

\item{$K$;\ Kasner (Vacuum) Solution} $[\cos(\theta),\sin(\theta),0,0,0,0,0,0,0]$ where $-\pi<\theta\leq\pi$.  This point is part of the Kasner surface discussed below.  

\end{description}

\subsubsection{Massless Scalar Field, $^0\Omega^{0,\pm}$}

\begin{description}
\item {${\cal K}$;\ Kasner (Massless Scalar Field) Solution} $[\sin(\varphi)\cos(\theta),\sin(\varphi)\sin(\theta),0,0,0,0,0,\cos(\varphi),0]$ where $-\pi<\theta\leq\pi$ and $-\frac{\pi}{2}\leq \varphi \leq \frac{\pi}{2}$. Or $\Sigma_+^2+\Sigma_-^2+\Psi^2=1$.  At this equilibrium point the deceleration parameter $q=2$.
The eigenvalues of the linearization restricted to the surface $G({\bf X})=0$ are ($\Omega_{\Lambda}$ eliminated) 
$$ -6, 0,0,2,2-4\Sigma_+,6-3\gamma,6+\sqrt{6}k\Psi,\{6\}$$
The zero eigenvalues correspond to the fact that this is a two-dimensional set of equilibria.  We note that this equilibrium set has at least one negative and one positive eigenvalue, and hence this point is always a saddle, even in the $\Omega_{\Lambda}=0$ invariant set.   However, in the invariant set $\Omega_{\lambda}=0$, a portion of this set becomes a local source.  

\end{description}

\subsubsection{Massive Scalar Field, $^0\Omega^{+,\pm}$}

\begin{description}
 
\item{$SF^0$;\ Robertson-Walker Scalar Field Solution (Power Law Inflationary Solution)}
$[0,0,0,0,0,0,0,-\frac{k\sqrt{6}}{6},1-\frac{k^2}{6}]$ At this equilibrium point the deceleration parameter $q=\frac{k^2}{2}-1$.  The eigenvalues of the linearization restricted to the surface $G({\bf X})=0$ are ($\Omega_{\Lambda}$ eliminated)  
$$ -k^2, k^2-4,k^2-3\gamma,\frac{1}{2}(k^2-6),\frac{1}{2}(k^2-6),\frac{1}{2}(k^2-6),\frac{1}{2}(k^2-2), \{k^2\}$$
We observe that this point has both negative and positive eigenvalues and that the point is not in the physical phase space for $k^2>6$.   In the $\Omega_{\Lambda}=0$ invariant set,
\begin{itemize}
\item if $k^2<2$ and $1\leq \gamma \leq 2$ then this point is stable.
\item if $2<k^2<3\gamma$ and $1\leq \gamma <4/3$ then this point has a 1 dimensional unstable manifold.  
\item if $3\gamma<k^2<4$ and $1\leq \gamma <4/3$ then this point has a 2 dimensional unstable manifold.
\item if $4<k^2<6$ and $1\leq \gamma <4/3$ then this point has a 3 dimensional unstable manifold.
\item if $2<k^2<4$ and $4/3\leq \gamma \leq 2$ then this point has a 1 dimensional unstable manifold.  
\item if $4<k^2<3\gamma$ and $4/3\leq \gamma \leq 2$ then this point has a 2 dimensional unstable manifold.
\item if $3\gamma<k^2<6$ and $4/3\leq \gamma \leq 2$ then this point has a 3 dimensional unstable manifold.
\end{itemize}
We therefore conclude that for $k^2<2$, the power law inflationary solution is an attractor in the $\Omega_{\Lambda}=0$ invariant set.

\item{$SF^{II}$;\ Bianchi II Scalar Field Solution}
$[2\frac{(k^2-2)}{k^2+16},0,\sqrt{\frac{3(2-k^2)(k^2-8)}{(k^2+16)^2}},0,0,0,0
-\frac{3\sqrt{6}k}{k^2+16},36\frac{8-k^2}{(k^2+16)^2}]$.  At this equilibrium point the deceleration parameter $q=8\frac{k^2-2}{k^2+16}$.  This point only exists for parameter values in the range of $2\leq k^2\leq 8$. The eigenvalues of the linearization restricted to the surface $G({\bf X})=0$ are ($\Omega_{\Lambda}$ eliminated)
\begin{eqnarray*}
& -18\frac{k^2}{k^2+16},6\frac{k^2-8}{k^2+16},6\frac{k^2-8}{k^2+16},2\frac{7k^2-32}{k^2+16},\\
& 3\frac{6k^2-16\gamma-\gamma k^2}{k^2+16},\frac{3}{k^2+16}\left[(k^2-8)\pm\sqrt{(k^2-8)^2+12(k^2-8)(k^2-2)}\right],\{18\frac{k^2}{k^2+16}\}
\end{eqnarray*}
We observe that this point has both negative and positive eigenvalues and therefore this point will always be a saddlepoint. In the $\Omega_{\Lambda}=0$ invariant set
\begin{itemize}
\item if $1\leq \gamma < 4/3$ and $2<k^2<16\gamma/(6-\gamma)$ then this point is stable. 
\item if $1\leq \gamma < 4/3$ and $16\gamma/(6-\gamma)<k^2<32/7$ then this point has a one dimensional unstable manifold. 
\item if $1\leq \gamma < 4/3$ and $32/7<k^2$ then this point has a two dimensional unstable manifold. 
\item if $4/3 < \gamma < 2$ and $2<k^2<32/7$ then this point is stable. 
\item if $4/3 < \gamma < 2$ and $32/7<k^2<16\gamma/(6-\gamma)$ then this point has a one dimensional unstable manifold. 
\item if $4/3 < \gamma < 2$ and $16\gamma/(6-\gamma)<k^2$ then this point has a two dimensional unstable manifold.
\end{itemize}

\item{$RSF^0$;\ Robertson Walker Radiation-Scalar Field Scaling Solution}
$[0,0,0,0,0,0,\frac{k^2-4}{k^2},-\frac{2\sqrt{6}}{3k},\frac{4}{3k^2}]$. At this equilibrium point the deceleration parameter $q=1$.  The eigenvalues of the linearization restricted to the surface $G({\bf X})=0$ are ($\Omega_{\Lambda}$ eliminated)  
$$ -4,-1,-1,4-3\gamma,1,-\frac{1}{2k}(k\pm\sqrt{k^2+16(4-k^2)}),\{4\}$$
This point exists in the set $U^+\cup U^0$ only if $k^2\geq 4$.  We observe that this point has both negative and positive eigenvalues  and therefore this point will always be a saddlepoint.
In the $\Omega_{\Lambda}=0$ invariant set
\begin{itemize}
\item if $1\leq \gamma <4/3$ then this point has a 2 dimensional unstable manifold.
\item if $4/3< \gamma \leq 2$ then this point has a 1 dimensional unstable manifold.
\end{itemize}

\item{$RSF^{II}$;\ Bianchi II Radiation-Scalar Field Scaling Solution}
$[\frac{1}{4},0,\frac{1}{4},0,0,0,\frac{7k^2-32}{8k^2},-\frac{2\sqrt{6}}{3k},\frac{4}{3k^2}]$  At this equilibrium point the deceleration parameter $q=1$. The eigenvalues of the linearization restricted to the surface $G({\bf X})=0$ are ($\Omega_{\Lambda}$ eliminated) 
\begin{eqnarray*} & -4,-1,4-3\gamma,\{4\}\\
& -\frac{1}{4k^2}\left( (2k^{2})\pm \sqrt {(2k^2)^2-(2k^2)
\left [(23k^2-64)\pm \sqrt {(23k^2-64)^2-64k^2(7k^2-32)}\right ] }  \right)
\end{eqnarray*}
This point exists in the set $U^+\cup U^0$ only if $k^2\geq 32/7$.  This point is a saddle in the full phase space.  However, in the invariant set $\Omega_{\Lambda}=0$,
\begin{itemize}
\item if $1\leq \gamma < 4/3$ then this point has a one dimensional unstable manifold.
\item if $4/3 < \gamma < 2$ then this point is stable.
\end{itemize} 

\end{description}

\subsubsection{Perfect Fluid Models, $^+\Omega^{0,0}$}

\begin{description}
 
\item{$F^0$;\ Robertson Walker Perfect Fluid Solution}
$[0,0,0,1,0,0,0,0,0]$ At this equilibrium point the deceleration parameter $q=\frac{1}{2}(3\gamma-2)$.  The eigenvalues of the linearization restricted to the surface $G({\bf X})=0$ are ($\Omega_{\Lambda}$ eliminated)  
$$-3\gamma, \frac{3}{2}(\gamma-2),\frac{3}{2}(\gamma-2),\frac{3}{2}(\gamma-2),3\gamma-4,\frac{1}{2}(3\gamma-2), 3\gamma,\{3\gamma\} $$
We observe that this point has both negative and positive eigenvalues and therefore, this point will always be a saddlepoint.  However, in the invariant set $\Omega_{\Lambda}=0$,
\begin{itemize}
\item if $1\leq \gamma < 4/3$ then this point has a 2 dimensional unstable manifold.
\item if $4/3 < \gamma < 2$ then this point has a 3 dimensional unstable manifold.
\end{itemize} 

\item{$F^{II}$;\ Bianchi II Perfect Fluid Solution}
$[\frac{1}{8}(3\gamma-2),0,\frac{1}{8}\sqrt{3(3\gamma-2)(2-\gamma)},\frac{3}{16}(6-\gamma),0,0,0,0,0]$
 At this equilibrium point the deceleration parameter $q=\frac{1}{2}(3\gamma-2)$.  The eigenvalues of the linearization restricted to the surface $G({\bf X})=0$ are ($\Omega_{\Lambda}$ eliminated)  
\begin{eqnarray*}
&-3\gamma,3\gamma-4, \frac{3}{2}(\gamma-2),\frac{3}{2}(\gamma-2), 3\gamma,\{3\gamma\},\\
 &-\frac{3}{8}\left( 2(2-\gamma)\pm\sqrt{[2(2-\gamma)]^2+2(2-\gamma)(\gamma-6)(3\gamma-2)} \right) \end{eqnarray*}
We observe that this point has both negative and positive eigenvalues and therefore, this point will always be a saddlepoint. However, in the invariant set $\Omega_{\Lambda}=0$,
\begin{itemize}
\item if $1\leq \gamma < 4/3$ then this point has a 1 dimensional unstable manifold.
\item if $4/3 < \gamma < 2$ then this point has a 2 dimensional unstable manifold.
\end{itemize} 
\end{description}

\subsubsection{Perfect Fluid with a Massless Scalar Field, $^+\Omega^{0,\pm}$}

There are no equilibrium points in the interior of this invariant set. 

\subsubsection{Perfect Fluid with a Massive Scalar Field, $^+\Omega^{+,\pm}$}

\begin{description}
 
\item{$MSF^0$;\ Robertson Walker Matter-Scalar Field Scaling Solution}
$[0,0,0,1-\frac{3\gamma}{k^2},0,0,0,-\frac{\sqrt{6}\gamma}{2k},\frac{3\gamma(2-\gamma)}{2k^2}]$ At this equilibrium point the deceleration parameter $q=\frac{1}{2}(3\gamma-2)$.  The eigenvalues of the linearization restricted to the surface $G({\bf X})=0$ are ($\Omega_{\Lambda}$ eliminated) 
\begin{eqnarray*}
&-3\gamma, \frac{3}{2}(\gamma-2),\frac{3}{2}(\gamma-2),\frac{1}{2}(3\gamma-2),3\gamma-4, \{3\gamma\},\\
&\frac{3}{4}\left( (\gamma-2)\pm \sqrt{(\gamma-2)^2+8\gamma(\gamma-2)(1-\frac{3\gamma}{k^2})}\right)
\end{eqnarray*}
This point only exists in the physical phase space when $k^2\geq 3\gamma$. We observe that this point has both negative and positive eigenvalues and therefore, this point will always be a saddlepoint. However, in the invariant set $\Omega_{\Lambda}=0$,
\begin{itemize}
\item if $1\leq \gamma < 4/3$ then this point has a 1 dimensional unstable manifold.
\item if $4/3 < \gamma < 2$ then this point has a 2 dimensional unstable manifold.
\end{itemize}

\item{$MSF^{II}$;\ Bianchi II Matter-Scalar Field Scaling Solution}
$[\frac{1}{8}(3\gamma-2),0,\frac{1}{8}\sqrt{3(3\gamma-2)(2-\gamma)}, 1-\frac{3\gamma}{k^2}+\frac{1}{16}(2-3\gamma) ,0,0,0,-\frac{\sqrt{6}\gamma}{2k},\frac{3\gamma(2-\gamma)}{2k^2}]$ At this equilibrium point the deceleration parameter $q=\frac{1}{2}(3\gamma-2)$.  The eigenvalues of the linearization restricted to the surface $G({\bf X})=0$ are ($\Omega_{\Lambda}$ eliminated)
\begin{eqnarray*}
&-3\gamma, \frac{3}{2}(\gamma-2),3\gamma-4, \{3\gamma\},-\frac{3}{8}\left( A \pm \sqrt{A^2+\frac{1}{2}A(B\pm\sqrt{B^2+C})}\right)\\
&{\rm where}\\
&A=2(2-\gamma) \\
&B=24\gamma(\frac{2\gamma}{k^2}-1)+3(2-\gamma)^2\\
&C=64\gamma(3\gamma-2)(\frac{16\gamma}{k^2}+\gamma-6)
\end{eqnarray*}
This point only exists in the physical phase space when $1-\frac{3\gamma}{k^2}+\frac{1}{16}(2-3\gamma) >0$.  We observe that this point has both negative and positive eigenvalues and therefore, this point will always be a saddlepoint. 
However, in the invariant set $\Omega_{\Lambda}=0$,
\begin{itemize}
\item if $1\leq \gamma < 4/3$ then this point is stable.
\item if $4/3 < \gamma < 2$ then this point has a 1 dimensional unstable manifold.
\end{itemize}  
 \end{description}


\section{Observations}

   The early time behaviour of the brane-world cosmologies under consideration is no longer characterized by a Kasner type singularity.  It is found that the initial singularity is represented by the isotropic brane-world solution.  It is also observed that this initial singularity is very much dominated by the kinetic energy of the scalar field.  It is worthy to note that the Kasner solutions are not stable to the past in the brane-world scenario, however, they continue to have an unstable manifold having a large dimension, and are indeed stable to the past for General Relativistic cosmologies.  

 It is no surprise that the De Sitter solution represents the late time behaviour of any cosmology with a positive four dimensional cosmological constant $\Lambda$.  The early time behaviour of any cosmology with a positive four dimensional cosmological constant is asymptotic to a regime with an equivalent four dimensional cosmological constant equal to zero.  That is, the early time behaviour of cosmologies having a positive four-dimensional cosmological constant $\Omega_{\Lambda}>0$, can be determined by investigating the behaviour of the governing dynamical system restricted to the $\Omega_{\Lambda}=0$ invariant set.

  We observe that the late time behaviour in the $\Omega_{\Lambda}=0$ invariant set is dependent upon the curvature of the models under consideration. For example, the late time behaviour of the Bianchi II cosmological models is one of four possible behaviours/solutions.  If $k^2<2$ then the isotropic power-law inflationary solution is the stable attractor.  If $k^2>2$ then the late time behaviour is one in which the models fail to isotropize, and have either a scalar field source, a combination of a scalar field and an ordinary matter source, or a combination of a scalar field and a dark radiation source depending upon the values of $k$ and $\gamma$.  (See table \ref{table1}).  On the other hand, the late time behaviour of the Bianchi I cosmological models is also one of four possible solutions.  If $k^2<2$ then the isotropic power-law inflationary solution is the stable attractor.  If $k^2>2$ then the late time behaviour is one in which the models continue to isotropize, and have either a scalar field source, a combination of a scalar-field and ordinary matter source, or a combination of a scalar field and a dark radiation source depending upon the values of $k$ and $\gamma$.  (See table \ref{table1} for summary details).

For the most part, it becomes very obvious that the existence of the scalar field in these brane-world cosmologies affects the initial dynamics when compared to the perfect fluid models studied in \cite{Campos01a,Campos01b}. The scalar field also plays an important role in the future dynamics of those models with a zero cosmological constant, that is, the scalar field is a dominant feature in the future dynamics of both the Bianchi type II models and the Bianchi type I models.  Two solutions of particular interest, the radiation-scalar field scaling solutions ($RSF^{II}, RSF^{0}$) are shown to be stable for particular values of $\gamma$ and $k$.  Since neither of these solutions exist in general relativistic cosmologies, and since both of these cosmologies are due to bulk effects acting on the brane, we have the conclusion, that in general (contrary to popular belief), these brane-world scenario's are not necessarily asymptotic to general relativistic cosmologies to the future.

The qualitative analysis for the bifurcation values $\gamma=1$ and $\gamma=4/3$, and for the various values of $k$ has not been completed.  However, this analysis is not expected to yield any new observations.  Additional numerical calculations do not reveal anything unexpected.  See figures \ref{figure2}--\ref{figure6} for phase portraits.  Note, the analysis presented here also does not address the case in which the local energy density due to non-local effects from the free gravitational field is negative, $\Omega_{\cal U}<0$.  If  $\Omega_{\cal U}<0$, then these traditionally ever-expanding models could potentially recollapse due to the negativity of this dark radiation.   This analysis is presently under consideration in  \cite{vandenHoogen02b}.

\begin{table} \label{table1}
\caption{Sinks and Sources for the Bianchi type II models: Set represents the invariant set containing the basin of attraction/repulsion, and the Dimension is the dimension of the basin of attraction  }
\begin{tabular}{|c|c|c|c|c|}
\hline
Equilibrium Point & Sink/Source & Set & Dimension & Conditions  \\
\hline
\hline
$m$         & Source    & $\Omega_{\lambda}\not = 0$    & 8 & $\gamma>1$\\
$DS$        & Sink      & $\Omega_{\Lambda}\not =0$     & 8 & \\
$\cal K$    & Source    & $\Omega_{\lambda}=0,N^+$      & 7 & $\Sigma_+<1/2,\Psi<-\sqrt{6}/k$ \\
$SF^0$      & Sink      & $\Omega_{\Lambda}=0,N^+$      & 7 & $k^2<2$ \\
$SF^{II}$   & Sink      & $\Omega_{\Lambda}=0,N^+$      & 7 & $1\leq \gamma < 4/3, 2<k^2<16\gamma/(6-\gamma)$ and \\  
            &           &                               &   &     $4/3\leq \gamma < 2, 2<k^2<32/7$\\
$MSF^{II}$  & Sink      & $\Omega_{\Lambda}=0,N^+$      & 7 & $1\leq \gamma < 4/3, k^2>16\gamma/(6-\gamma)$\\
$RSF^{II}$  & Sink      & $\Omega_{\Lambda}=0,N^+$      & 7 & $4/3 < \gamma < 2$, $k^2>32/7$ \\
$R^{II}$    & Sink      & $\Omega_{\Lambda}=0,N^+$      & 7 & $4/3 < \gamma \leq 2$ \\
$RSF^0$     & Sink      & $\Omega_{\Lambda}=0,N^0$      & 6 & $4/3 < \gamma \leq 2, k^2>4$ \\
$SF^0$      & Sink      & $\Omega_{\Lambda}=0,N^0$      & 6 & $1\leq \gamma < 4/3, k^2<3\gamma$ and \\
            &           &                               &   & $4/3 < \gamma \leq 2, k^2<4$ \\
$MSF^{0}$   & Sink      & $\Omega_{\Lambda}=0,N^0$      & 6 & $1\leq \gamma < 4/3, k^2>3\gamma$\\
$R^{0}$     & Sink      & $\Omega_{\Lambda}=0,N^0$      & 6 & $4/3 < \gamma \leq 2$ \\
\hline
\end{tabular}
\end{table}


\begin{acknowledgments}
RJvdH is supported by research grants through Natural Sciences and Engineering Research Council of Canada and the University Council on Research at St. Francis Xavier University.  The work has been supported by the CICYT(Spain) grant BFM 2000-0018.  RJvdH wishes to thank the Departamento de F{\'i}sica Te{\'o}rica of the Universidad del Pa{\'i}s Vasco in Bilbao for their generous support and warm hospitality.
\end{acknowledgments}


\centerline{{\bf References}}


\section*{Appendix: Equilibrium Points in the Vacuum Invariant Set}\label{Appendix_V}

We define spherical coordinates as follows,
\begin{eqnarray}
\tilde \rho^2 &=& \Omega_{pf} + \Psi^2 + \Phi\nonumber\\
\tan^2 \tilde \theta & = & \frac{\Phi}{\Psi^2}\label{new-vars}\\
\cos^2 \tilde\phi & = & \frac{\Omega_{pf}}{\Omega_{pf}+\Psi^2 + \Phi}\nonumber
\end{eqnarray}

The evolution equations for $(\tilde \rho, \tilde \theta,\tilde \phi)$ are
\begin{eqnarray}
\tilde \rho ' & = & \tilde\rho\left( q + 1 - \frac{3\gamma}{2}\cos^2\tilde\phi-3\sin^2\tilde\phi\cos^2\tilde\theta\right),\\
\tilde\theta'&=& \sin\tilde\theta\left(3\cos\tilde\theta+\frac{\sqrt{6}}{2}k\tilde\rho\sin\tilde\phi\right),\\
\tilde\phi' & = & \sin\tilde\phi\cos\tilde\phi\left(\frac{3\gamma}{2}-3\cos^2\tilde\theta\right)
\end{eqnarray}
where 
$$q = 2\Sigma_{+}^2+2\Sigma_{-}^2 +\frac{3\gamma-2}{2}\tilde\rho^2\cos^2\tilde\phi - \Omega_{\Lambda}+\Omega_{\cal U} + \rho^2\sin^2\tilde\phi(2\cos^2\tilde\theta - \sin^2\tilde\theta) + \frac{3\gamma_{\lambda}-2}{2}\Omega_{\lambda}$$
and 
$$\gamma_{\lambda} = 2\gamma\cos^2\tilde\phi^2+4\sin^2\tilde\phi\cos^2\tilde\theta.$$
The constraint equation (\ref{constraint2}) becomes
\begin{equation}
\label{new-constraint}
\tilde\rho^2+\Omega_{\Lambda}+\Omega_{\lambda}+\Omega_{\cal U}+\Sigma_{+}^2+\Sigma_{-}^2+N^2 -1 =0
\end{equation}
and we have $0\leq \tilde \theta \leq \pi$ and $0\leq \tilde \phi\leq \pi/2$ since both $\Phi\geq 0$ and $ \Omega_{pf}\geq 0$.

If we define ${\bf Y} = (\tilde\rho,\tilde\theta,\tilde\phi,\Sigma_{+},\Sigma_{-},N,\Omega_{\lambda},\Omega_{\cal U})$ then we obtain a new dynamical system ${\bf Y}' = \tilde {\bf F}({\bf Y})$.  Where the dynamical equations for $\Sigma_{+},\Sigma_{-},N,\Omega_{\lambda},\Omega_{\cal U}$ are the same as those in equation (\ref{DS}) with the change of variables (\ref{new-vars}), and $\Omega_{\Lambda}$ is given by solving equation (\ref{new-constraint}) for $\Omega_{\Lambda}$.  Note, we indicate the eigenvalue corresponding to the $\Omega_{\Lambda}$ direction with brackets $\{\ \}$.

The geometrical interpretation of the analysis of this new system requires some care, as each equilibrium point will have a number of different coordinatizations in this new coordinate system.  The dynamics in the rectangular coordinate system $(\Omega_{pf},\Psi,\Phi)$ can be obtained by determining the dynamics along invariant rays or directions (the intersections of invariant planes $(\tilde\theta'=0)$ with invariant cones $(\tilde\phi'=0)$) in the spherical coordinate system.  We quickly observe here that $\tilde\theta=\pi/2$, $\tilde\phi=\pi/2$ does not correspond to an invariant direction of the dynamical system ${\bf Y}' = \tilde {\bf F}({\bf Y})$.

\begin{figure}[h]
\caption{Behaviour of the equilibrium points in the $\tilde\phi - \tilde\theta$ plane.}\label{figure1}
\begin{picture}(300,300)

\put(250,50){\vector(-1,0){197}}
\put(50,50){\vector(0,1){97}}
\put(250,50){\vector(0,1){97}}
\put(50,150){\vector(1,0){197}}
\put(50,250){\vector(0,-1){97}}
\put(250,250){\vector(-1,0){197}}
\put(250,250){\vector(0,-1){97}}

\put(250,50){\vector(-1,0){10}}
\put(50,50){\vector(0,1){10}}
\put(250,50){\vector(0,1){10}}
\put(50,150){\vector(1,0){10}}
\put(50,250){\vector(0,-1){10}}
\put(250,250){\vector(-1,0){10}}
\put(250,250){\vector(0,-1){10}}

\put(50,50){\circle*{5}}
\put(50,150){\circle*{5}}
\put(50,250){\circle*{5}}
\put(250,50){\circle*{5}}
\put(250,150){\circle*{5}}
\put(250,250){\circle*{5}}

\put(250,50){\line(1,0){20}}
\put(50,250){\line(0,1){20}}

\put(250,35){\makebox(10,0){\large $ \frac{\pi}{2} $ }}
\put(35,150){\makebox(10,0){\large $ \frac{\pi}{2} $ }}
\put(35,250){\makebox(10,0){\large $ \pi $ }}

\put(280,55){\makebox(10,0){\large $ \tilde\phi $ }}
\put(50,280){\makebox(5,0){\large $ \tilde\theta$ }}
\put(40,35){\makebox(10,0){\large $ 0$ }}

\end{picture}
\end{figure}
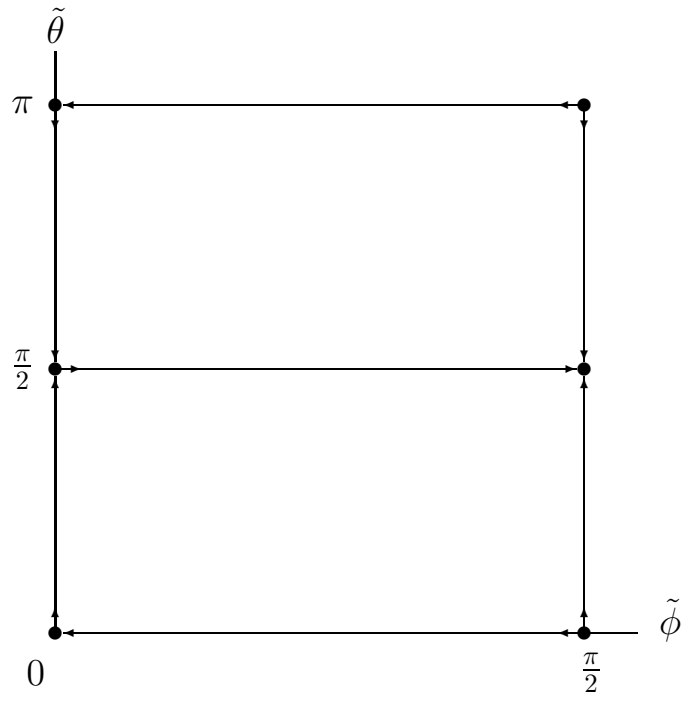

\begin{figure}[t]
\caption{Qualitative behaviour of trajectories when $\gamma=1$, $k=1$ and $k=2$.}\label{figure2}
\includegraphics[width=7cm]{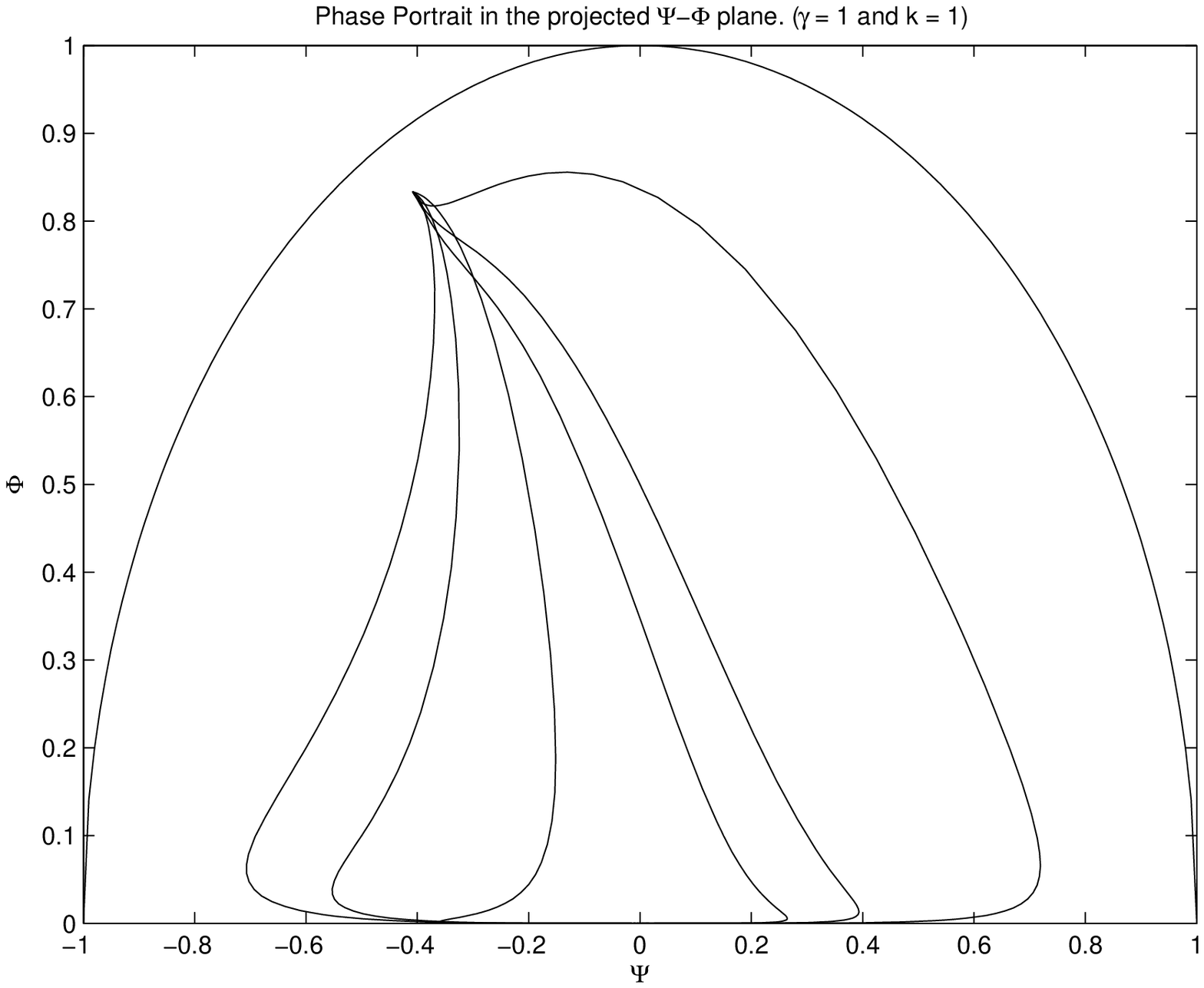}\qquad\includegraphics[width=7cm]{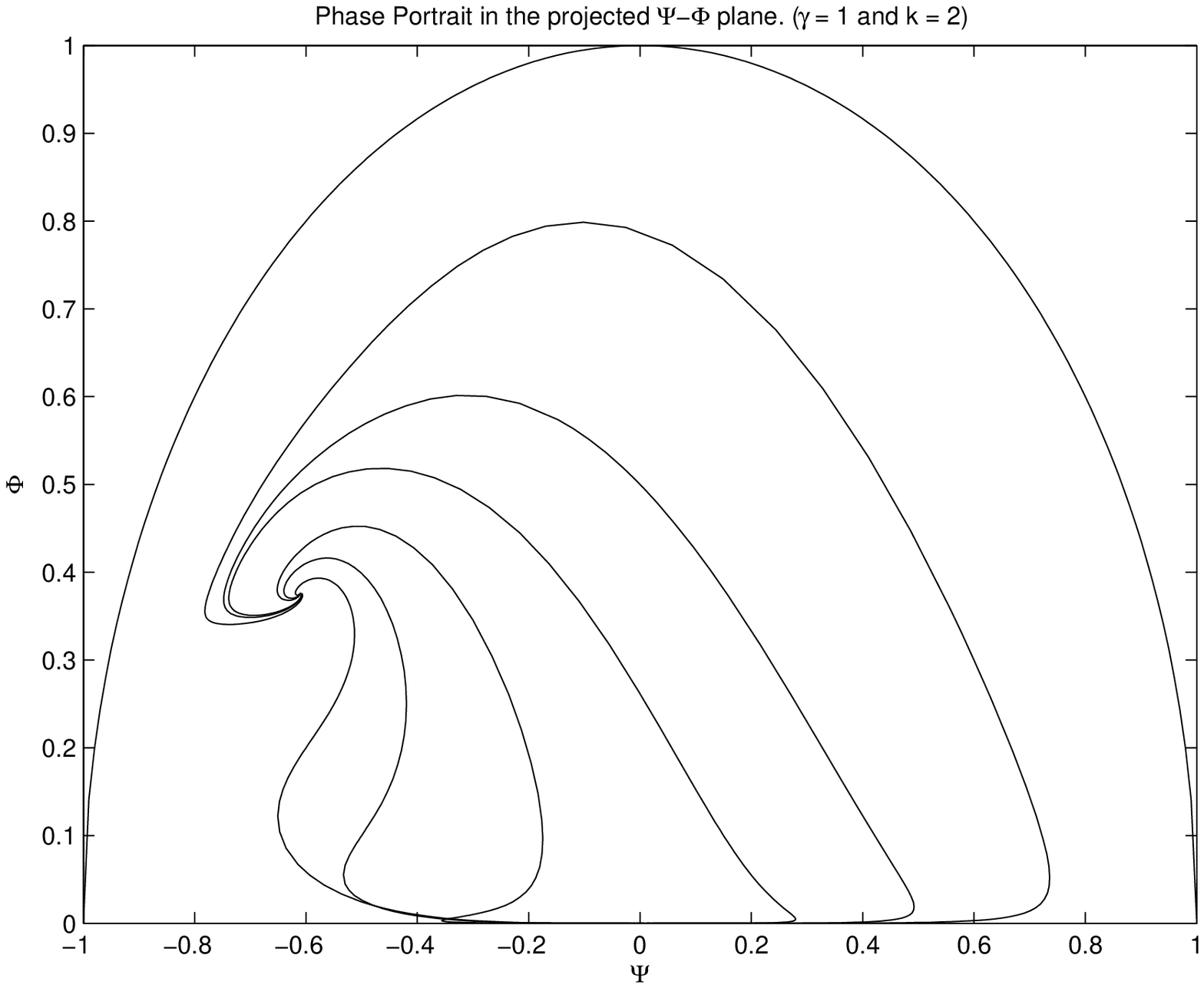}

\includegraphics[width=7cm]{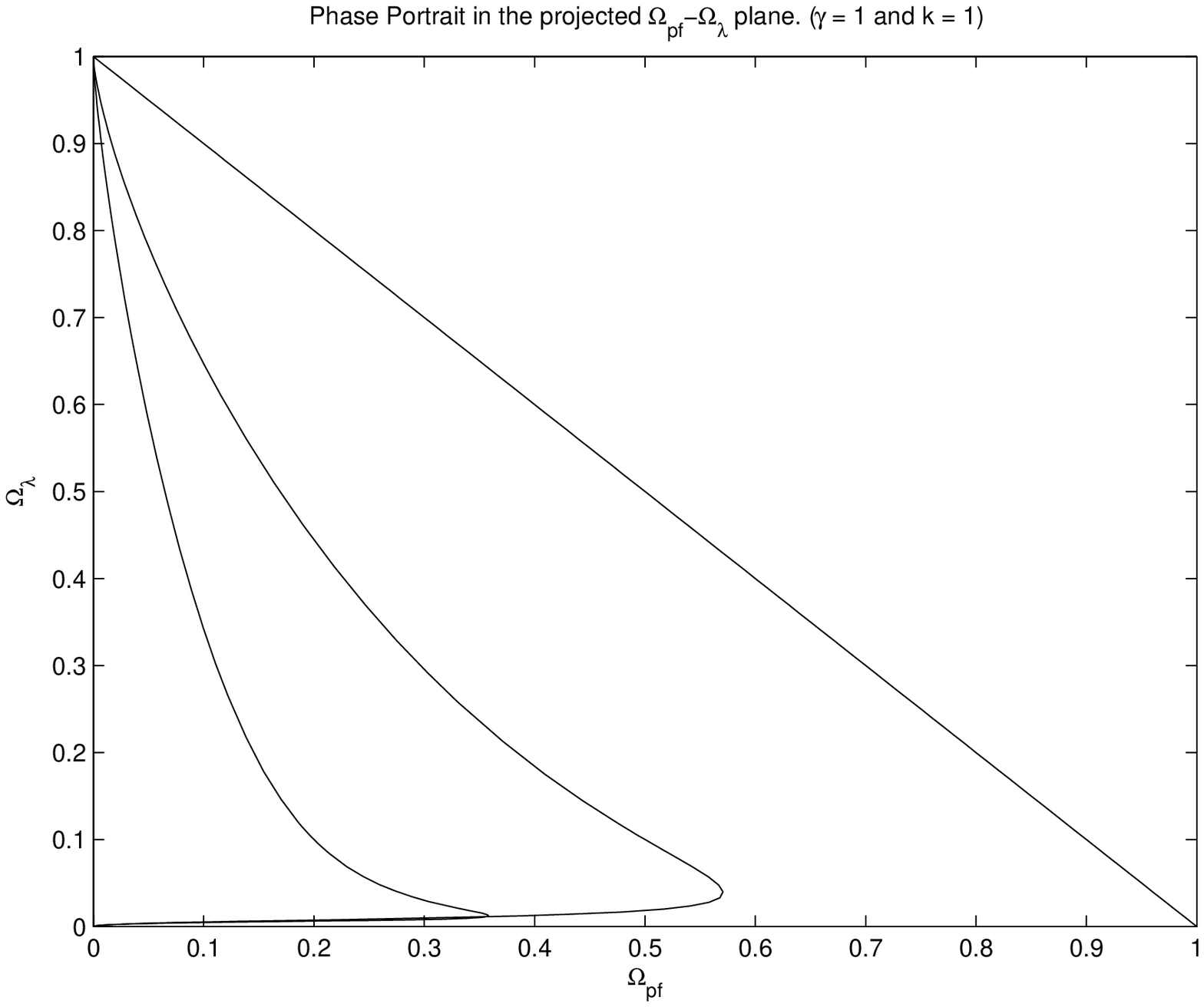}\qquad\includegraphics[width=7cm]{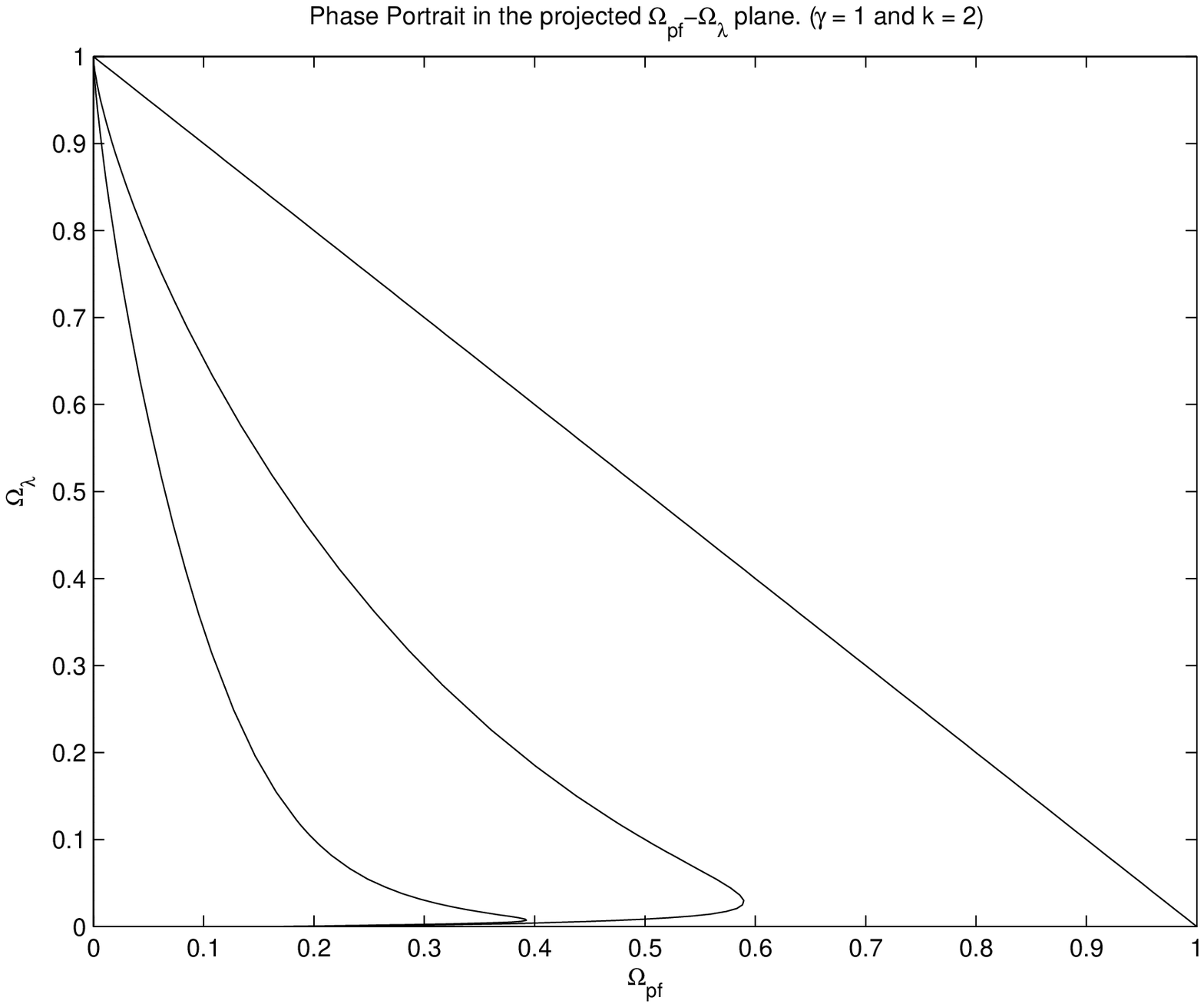}

\includegraphics[width=7cm]{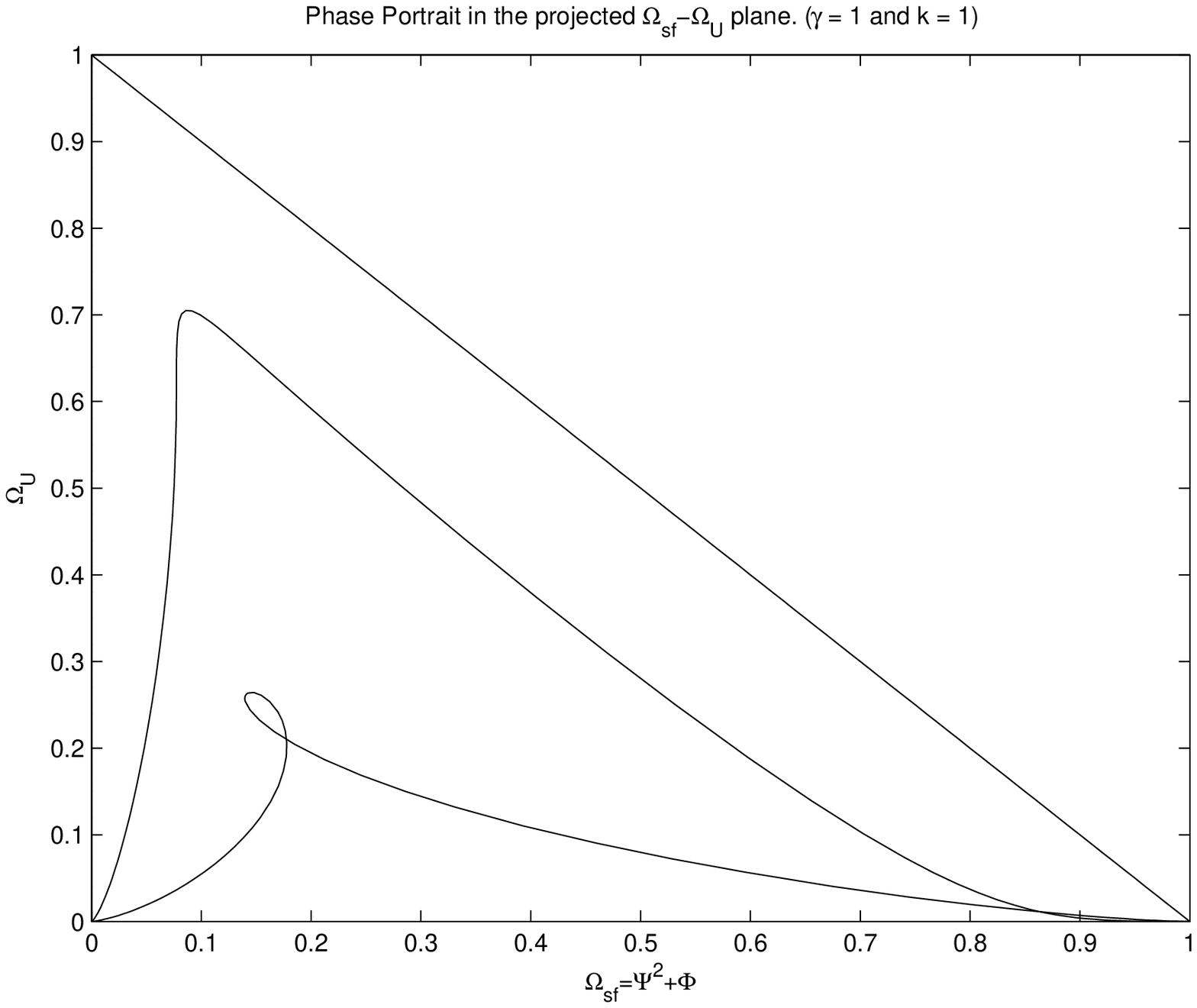}\qquad\includegraphics[width=7cm]{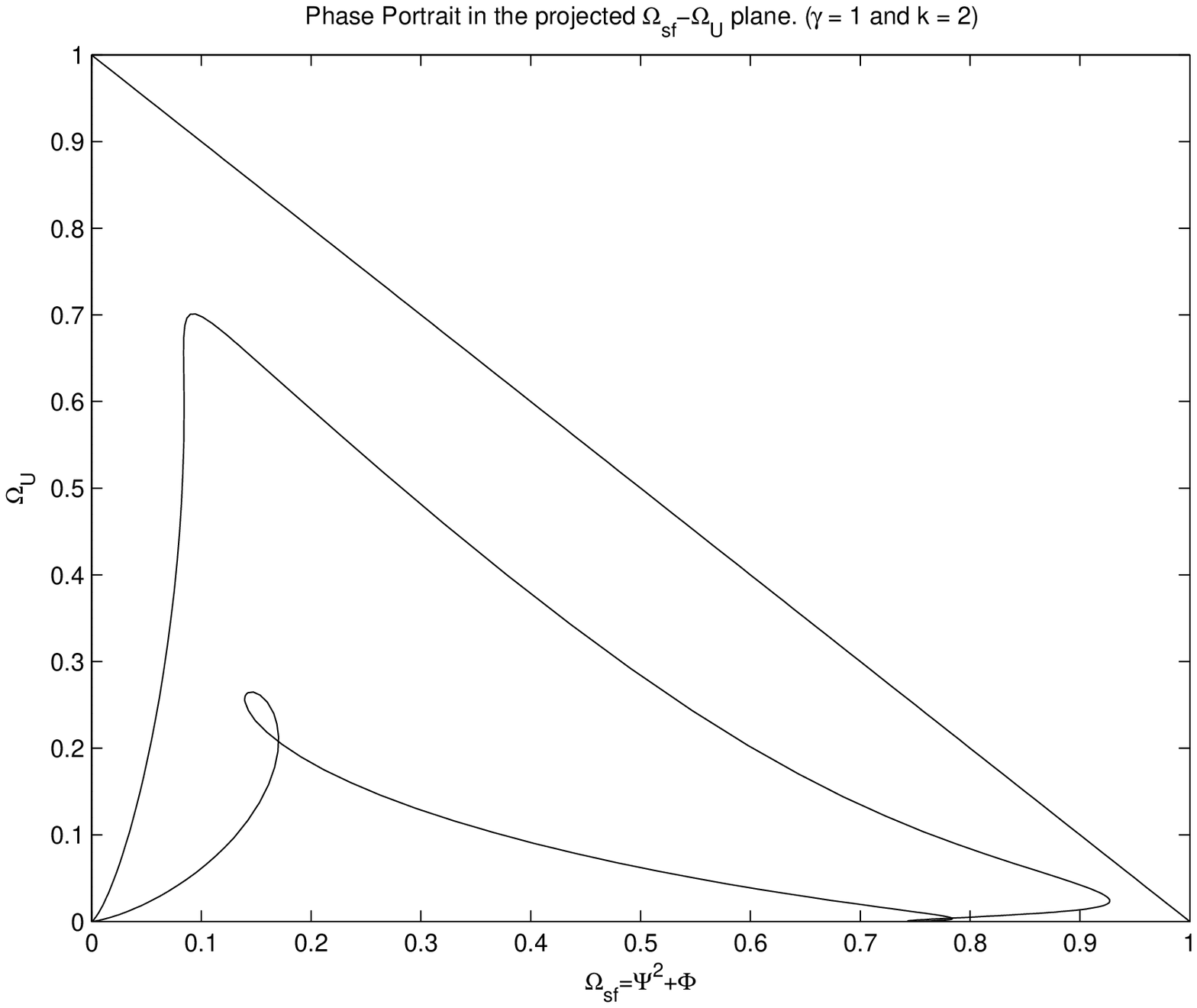}
\end{figure}

\begin{figure}[t]
\caption{Qualitative behaviour of trajectories when $\gamma=7/6<4/3$, $k=1$ and $k=2$.}\label{figure3}
\includegraphics[width=7cm]{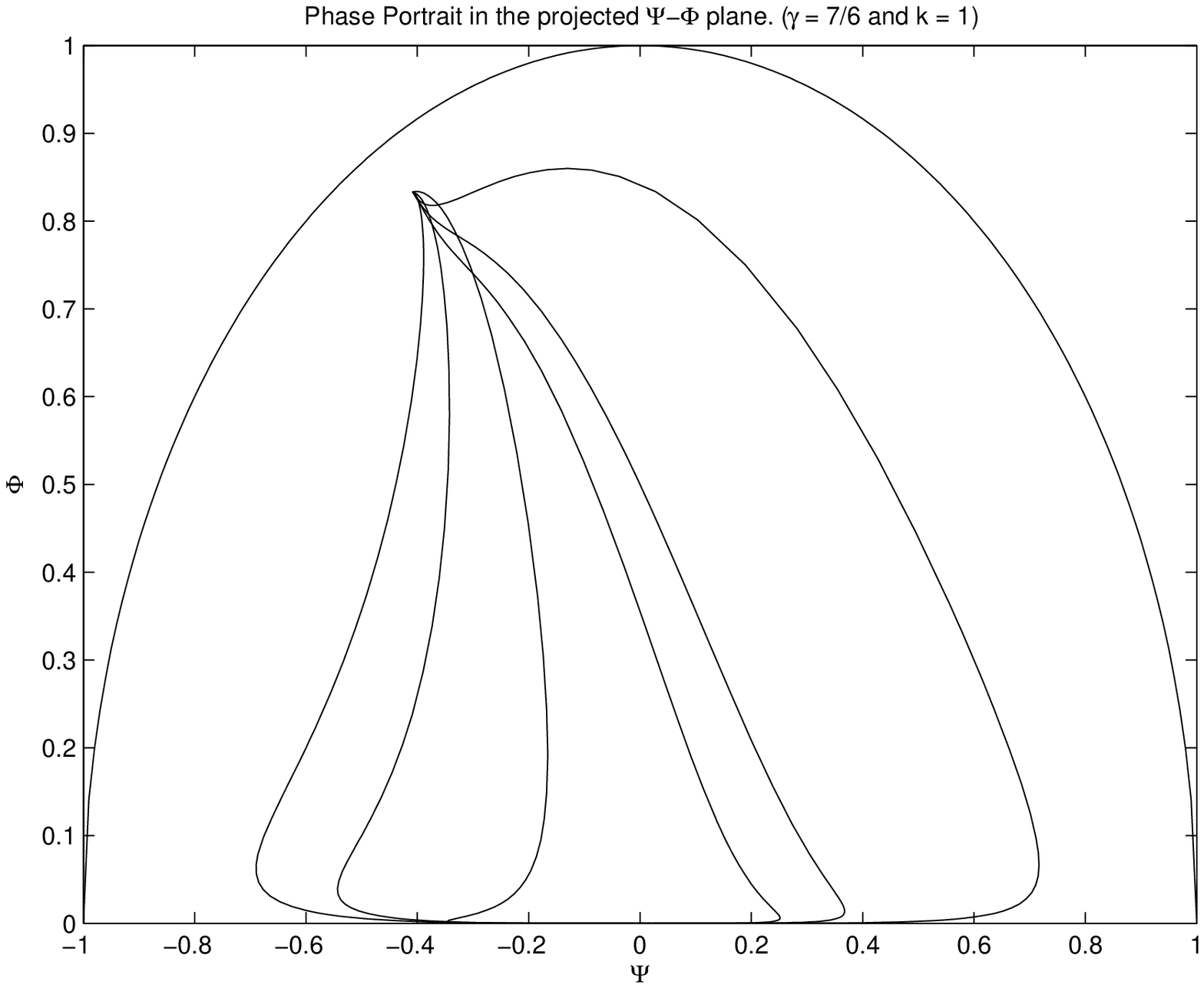}\qquad\includegraphics[width=7cm]{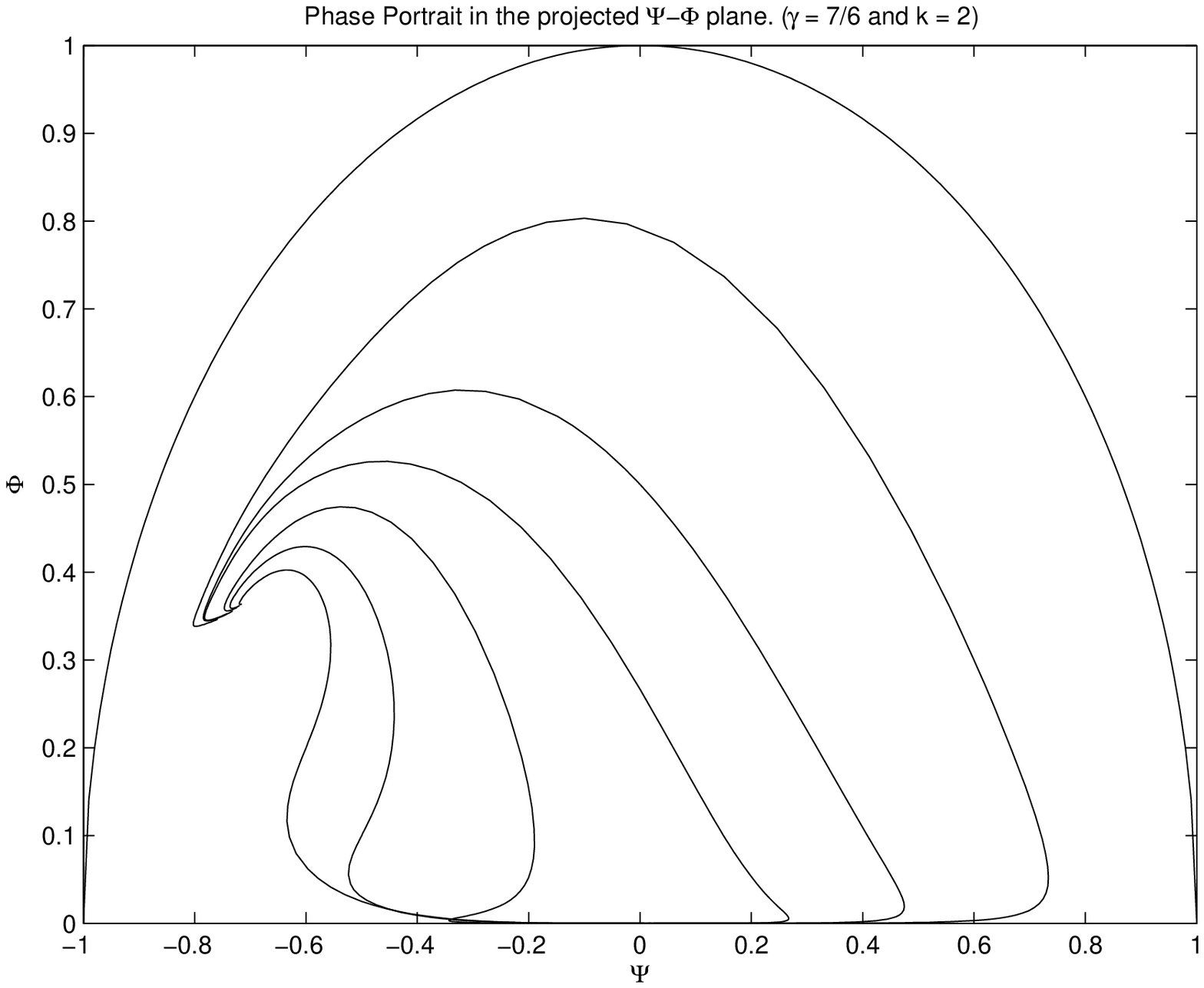}

\includegraphics[width=7cm]{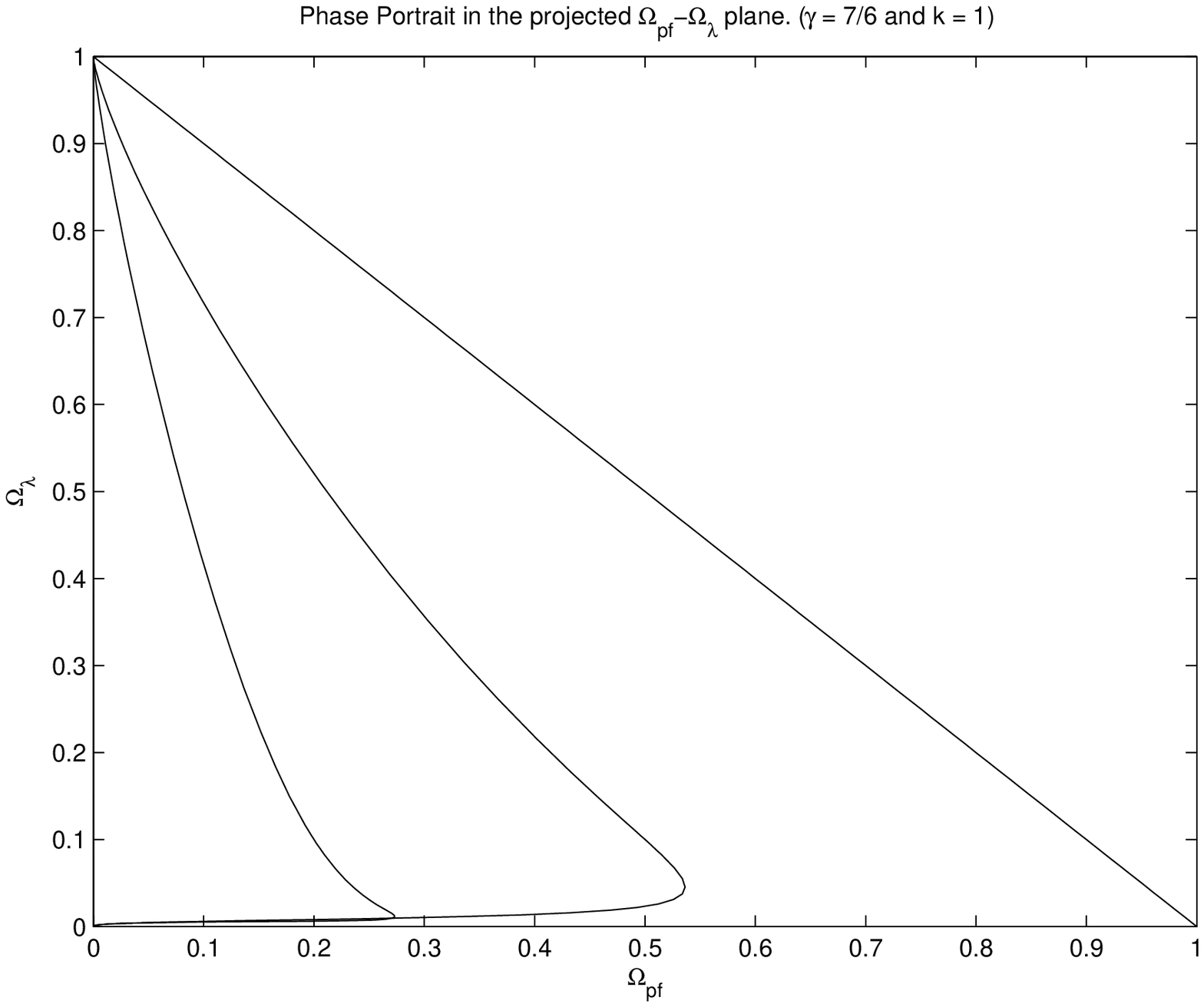}\qquad\includegraphics[width=7cm]{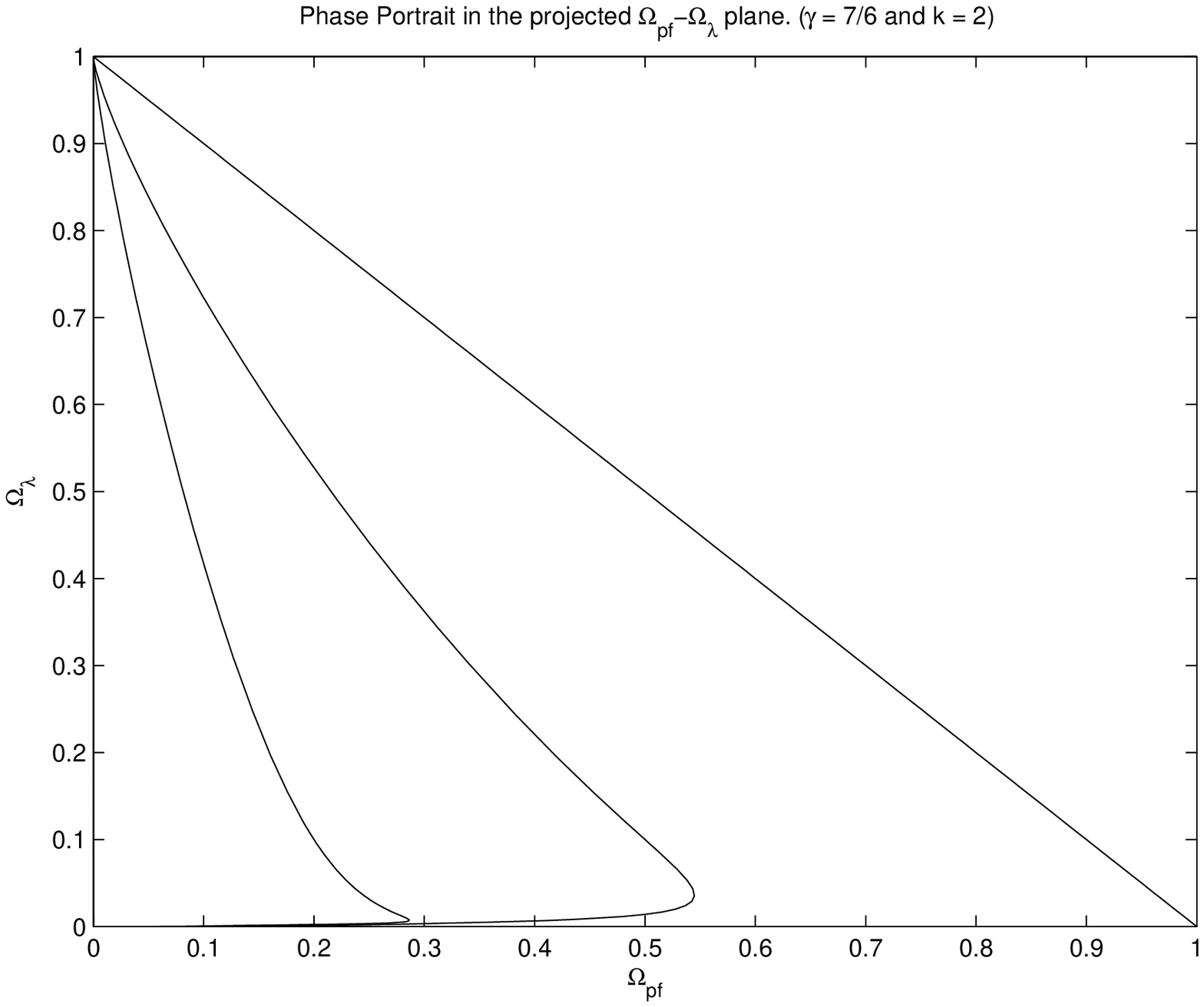}

\includegraphics[width=7cm]{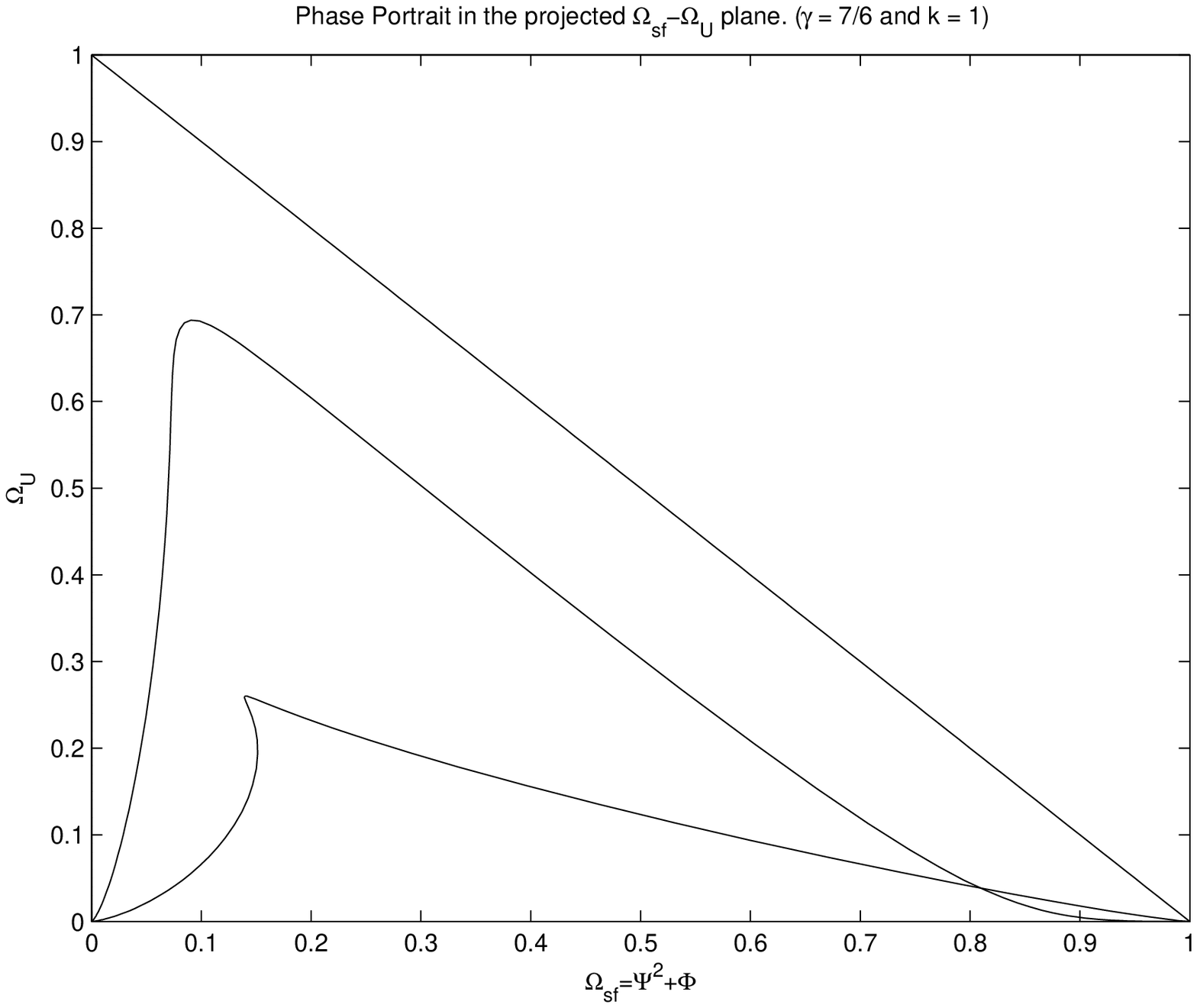}\qquad\includegraphics[width=7cm]{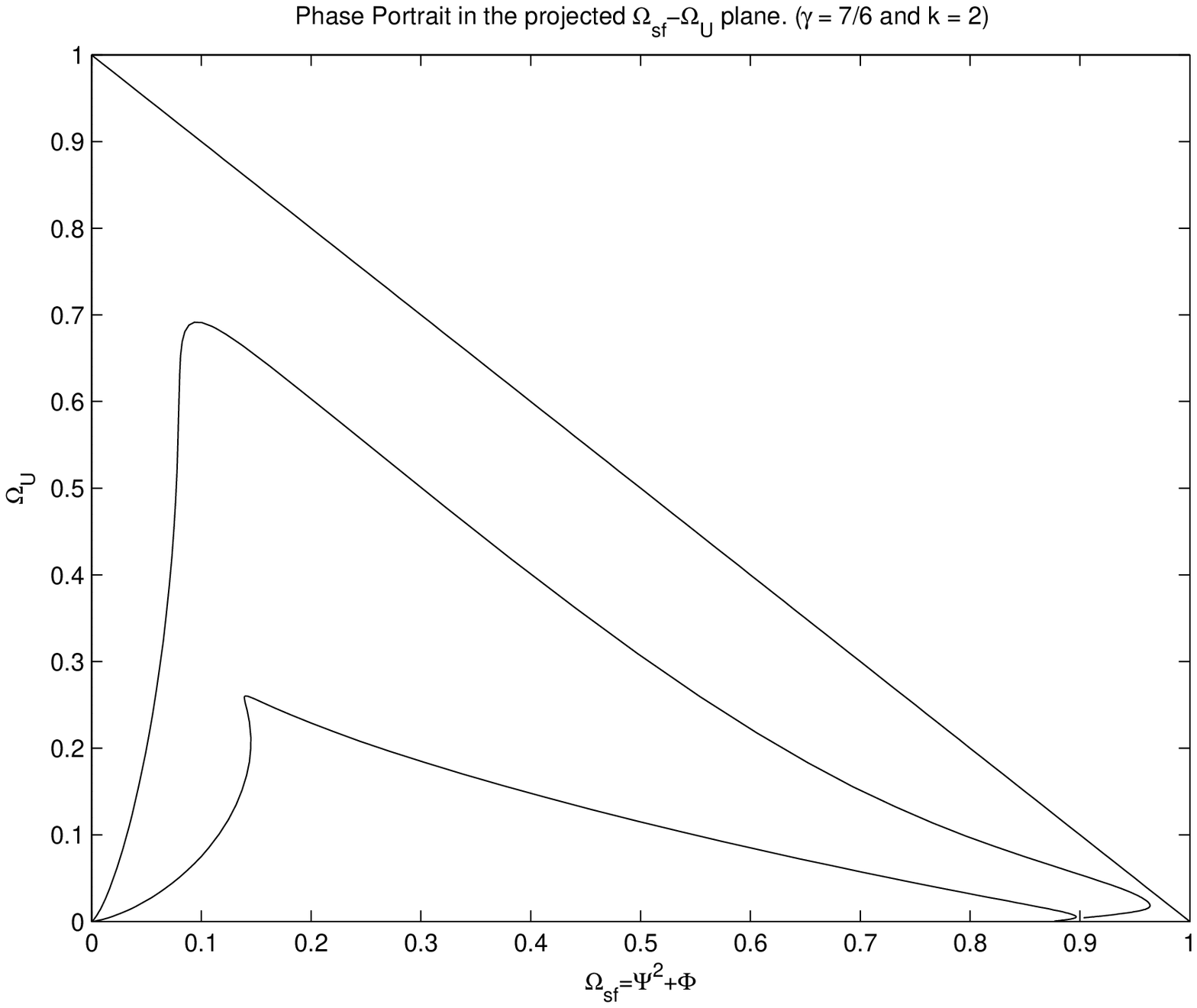}
\end{figure}

\begin{figure}[t]
\caption{Qualitative behaviour of trajectories when $\gamma=4/3$, $k=1$ and $k=2$.}\label{figure4}
\includegraphics[width=7cm]{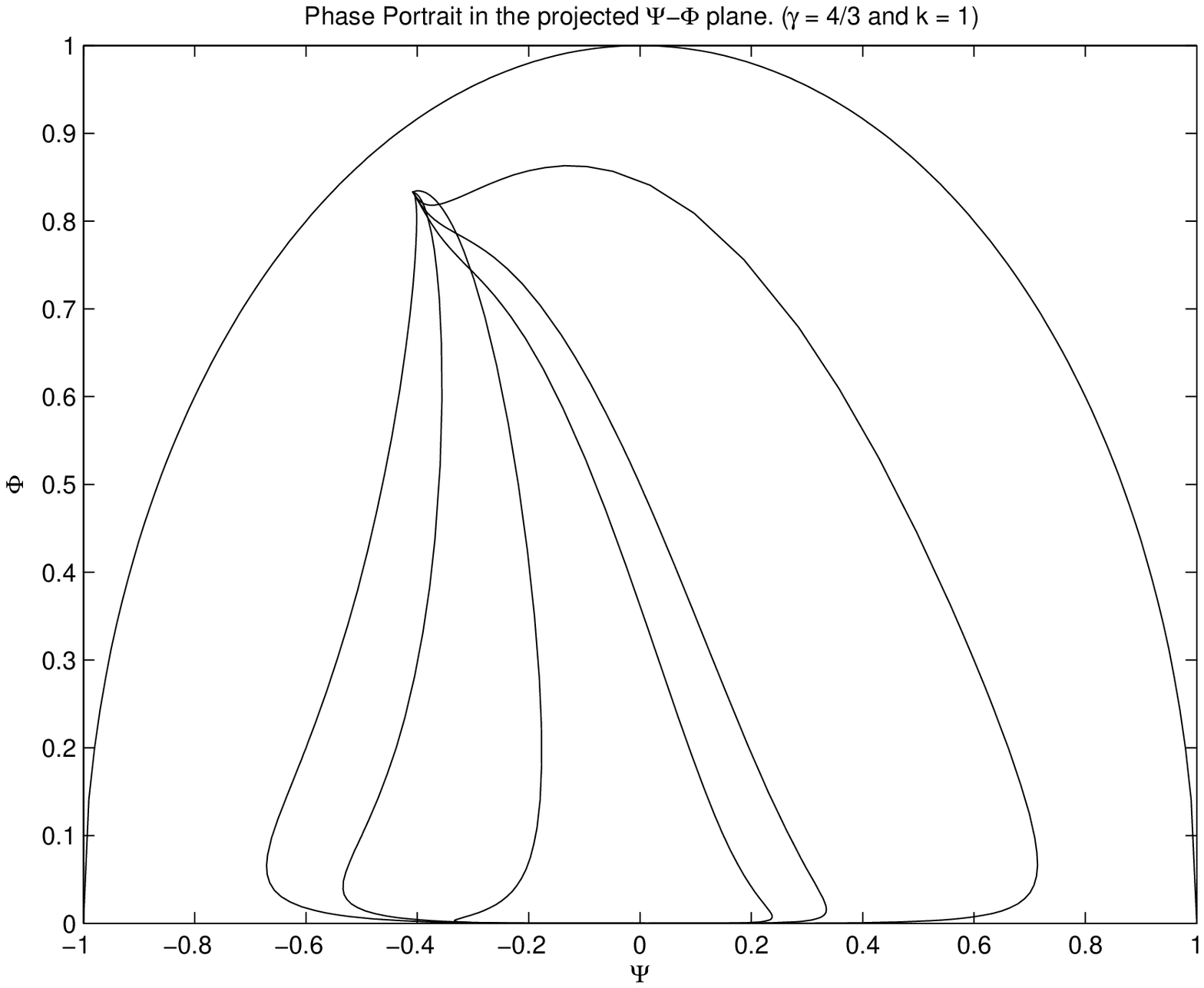}\qquad\includegraphics[width=7cm]{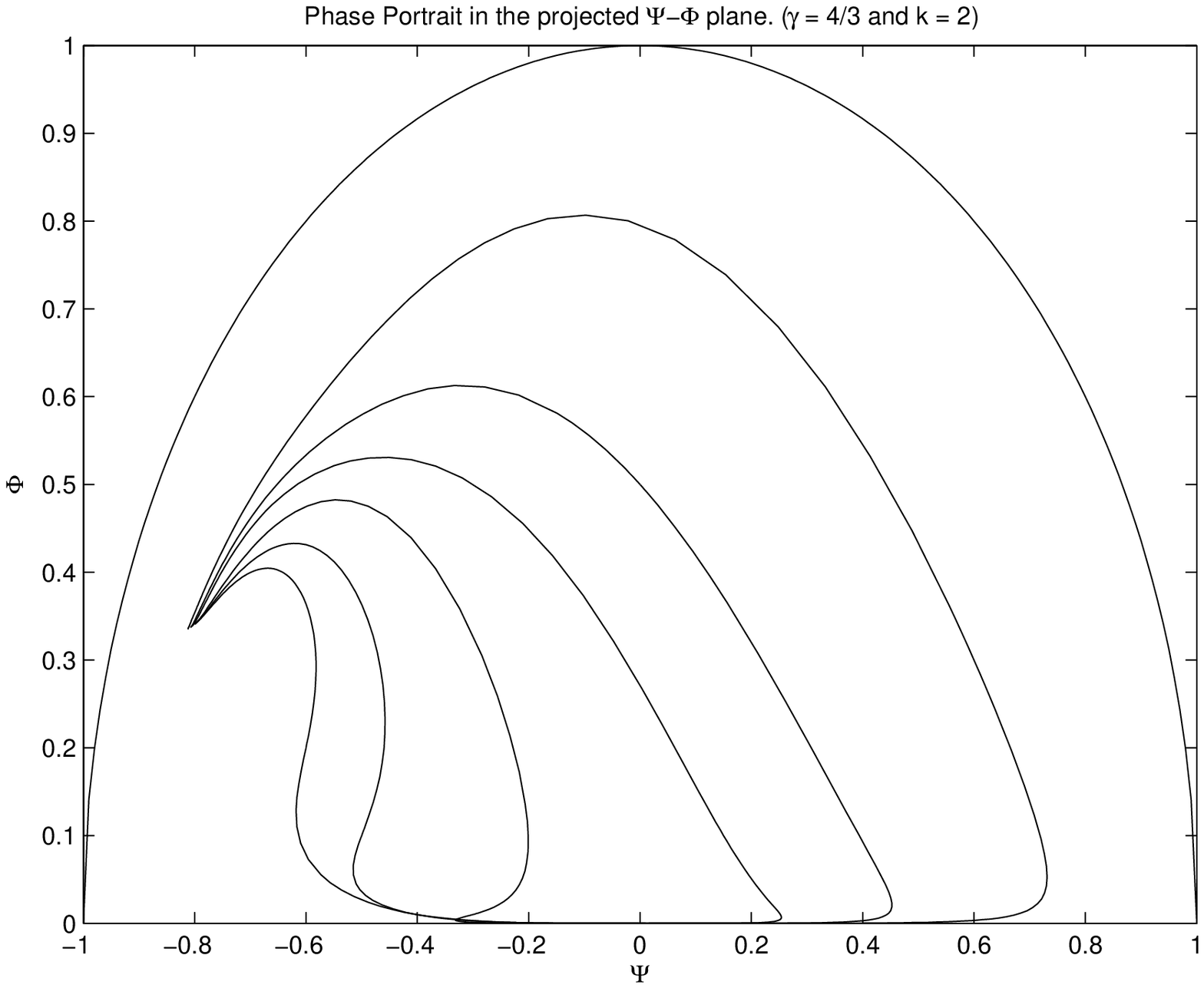}

\includegraphics[width=7cm]{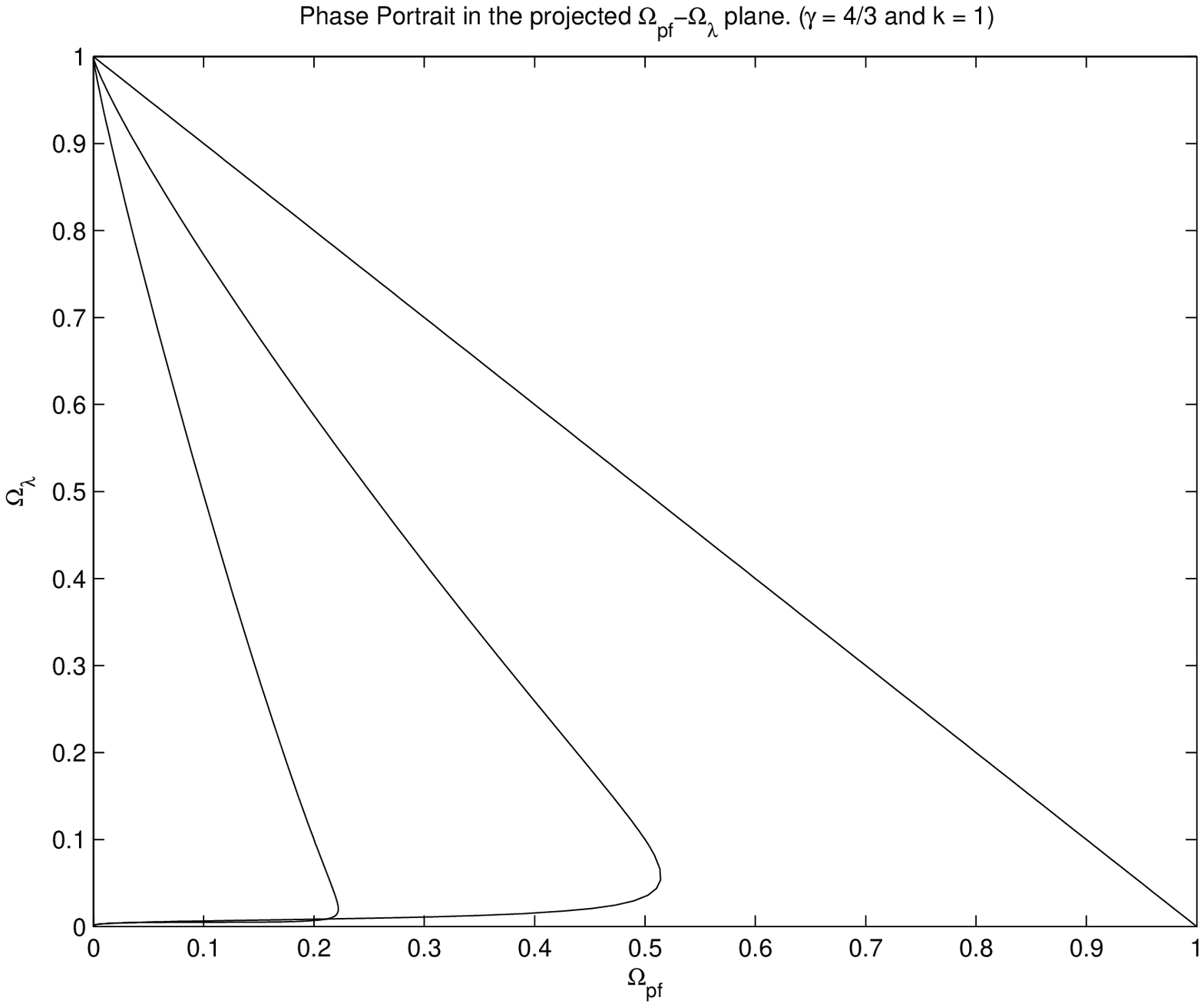}\qquad\includegraphics[width=7cm]{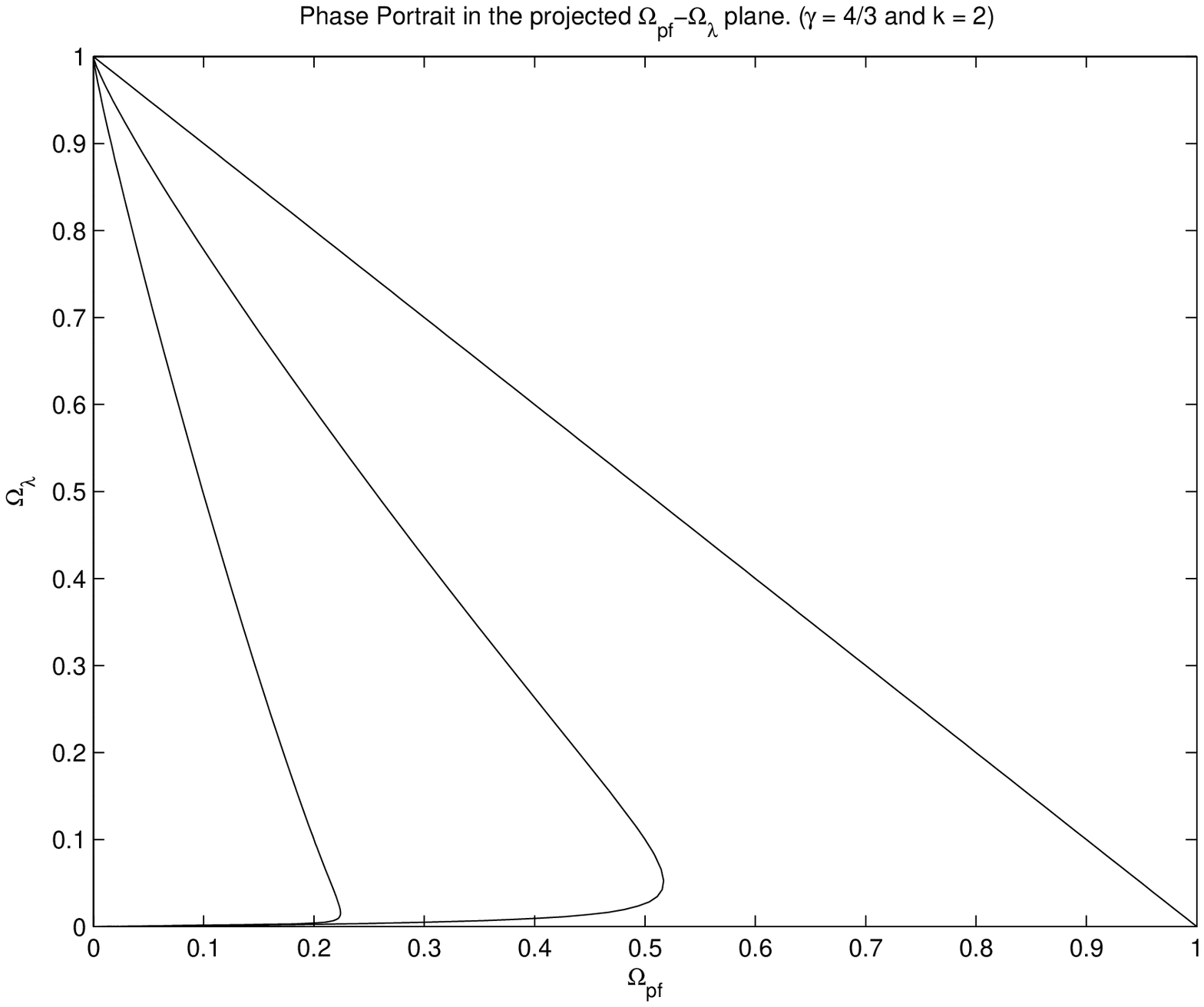}

\includegraphics[width=7cm]{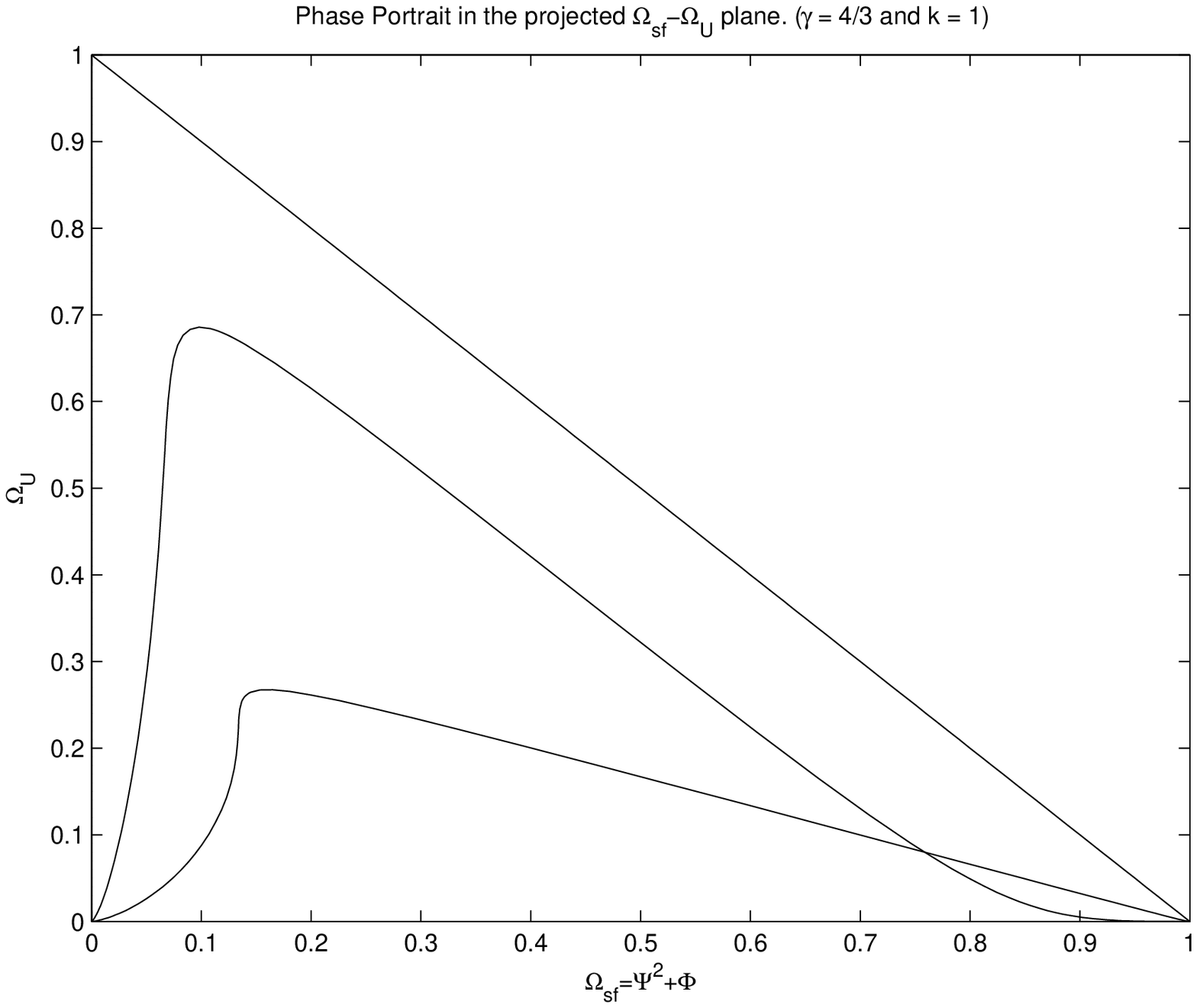}\qquad\includegraphics[width=7cm]{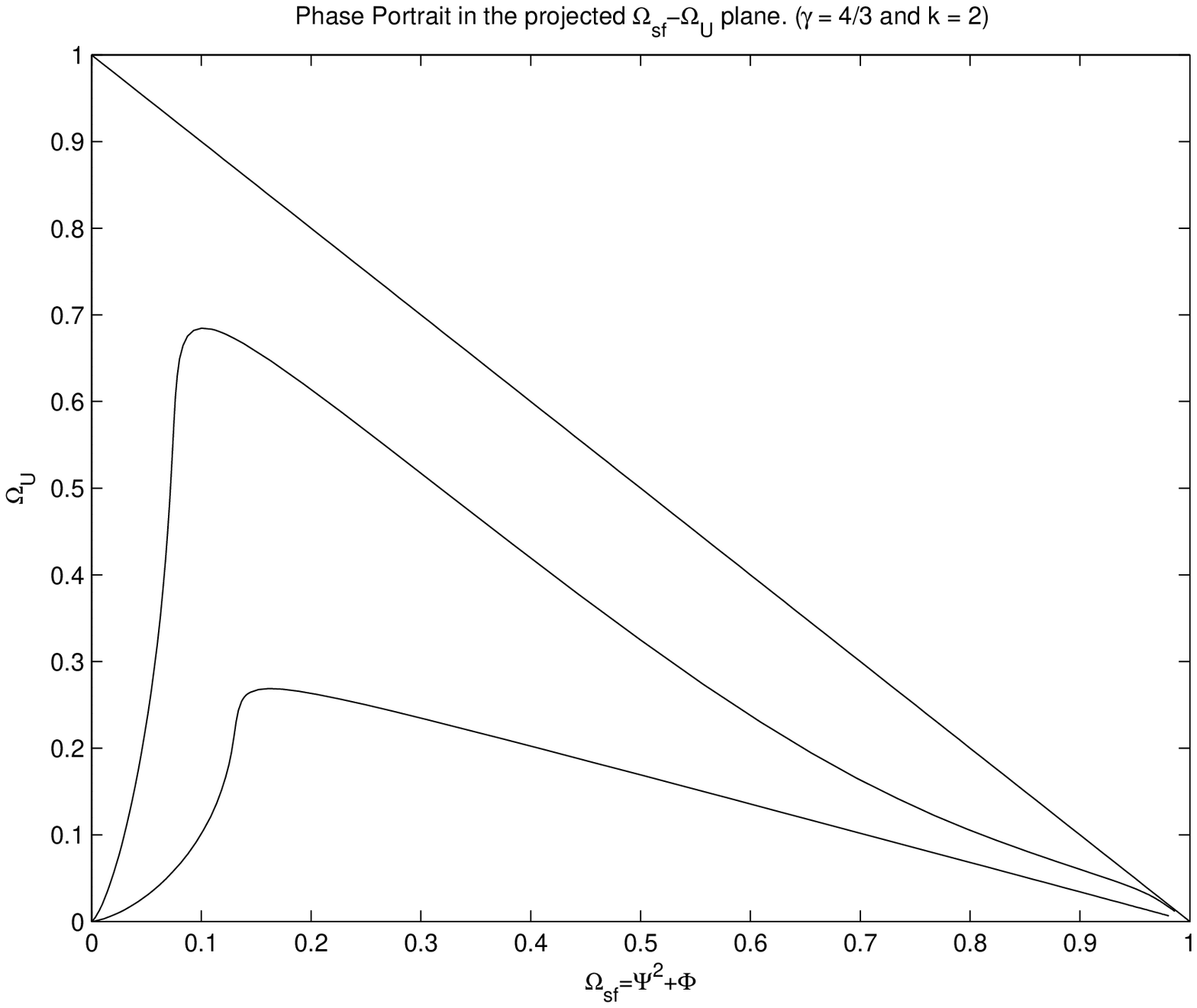}
\end{figure}

\begin{figure}[t]
\caption{Qualitative behaviour of trajectories when $\gamma=5/3>4/3$, $k=1$ and $k=2$.}\label{figure5}
\includegraphics[width=7cm]{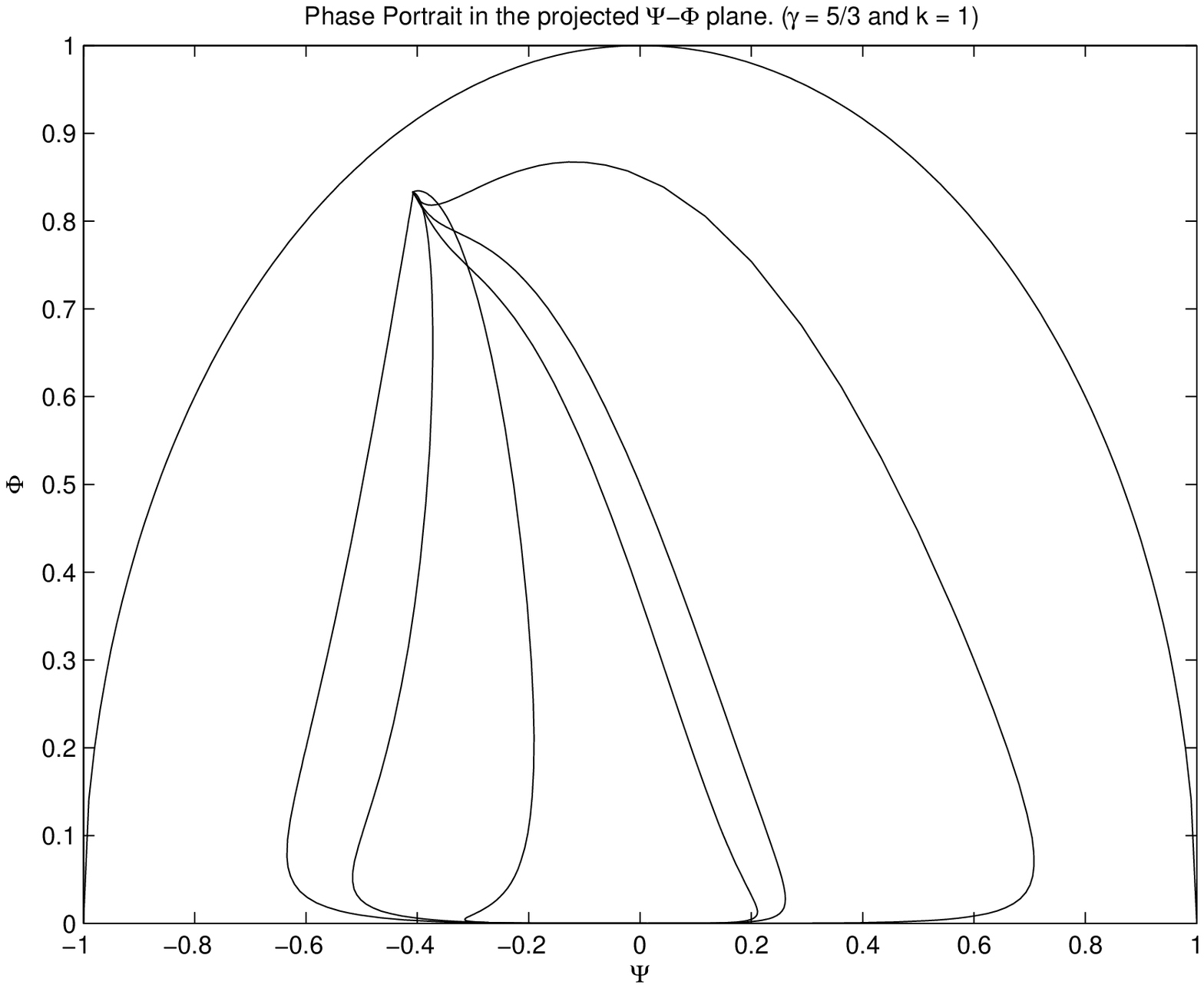}\qquad\includegraphics[width=7cm]{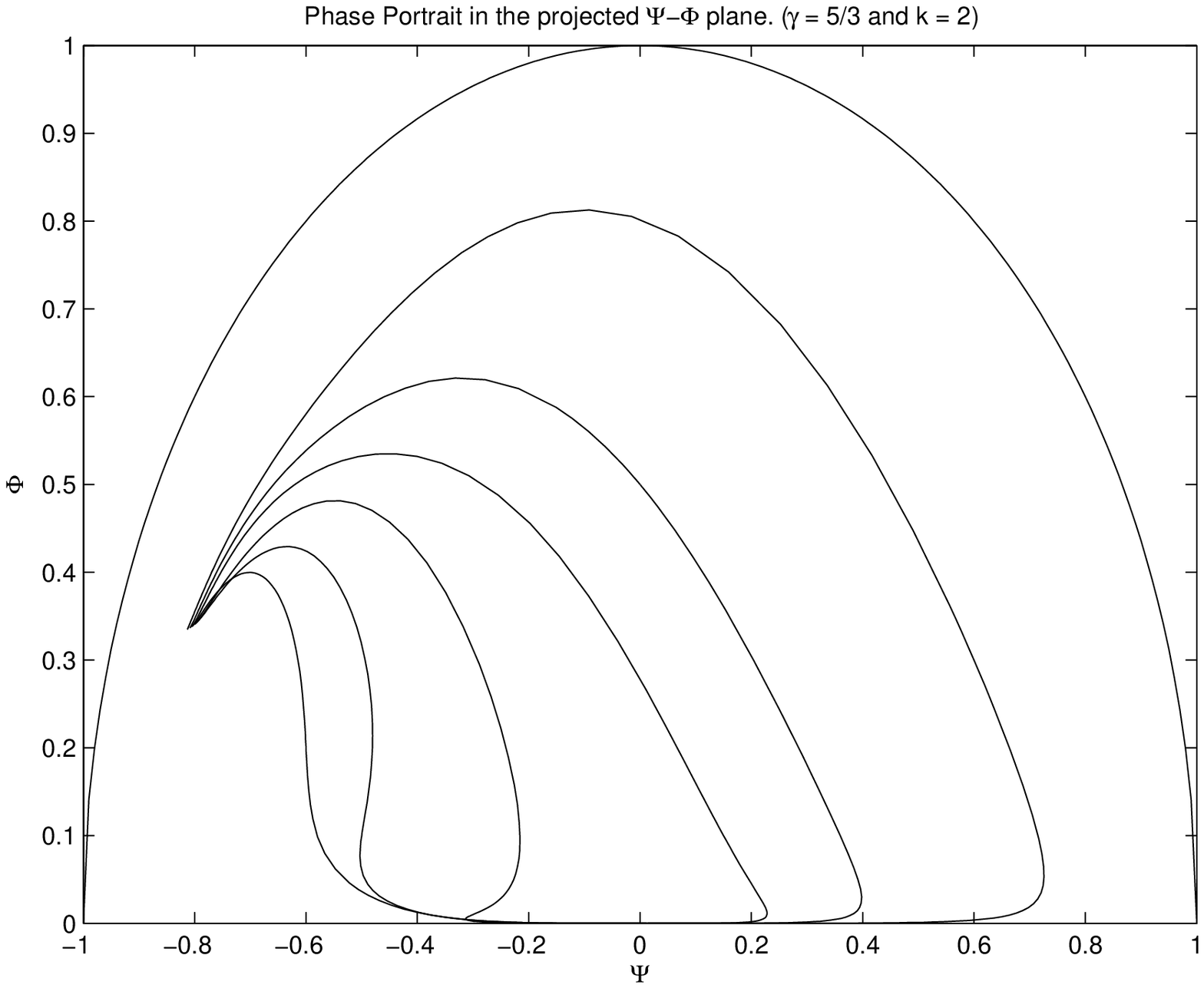}

\includegraphics[width=7cm]{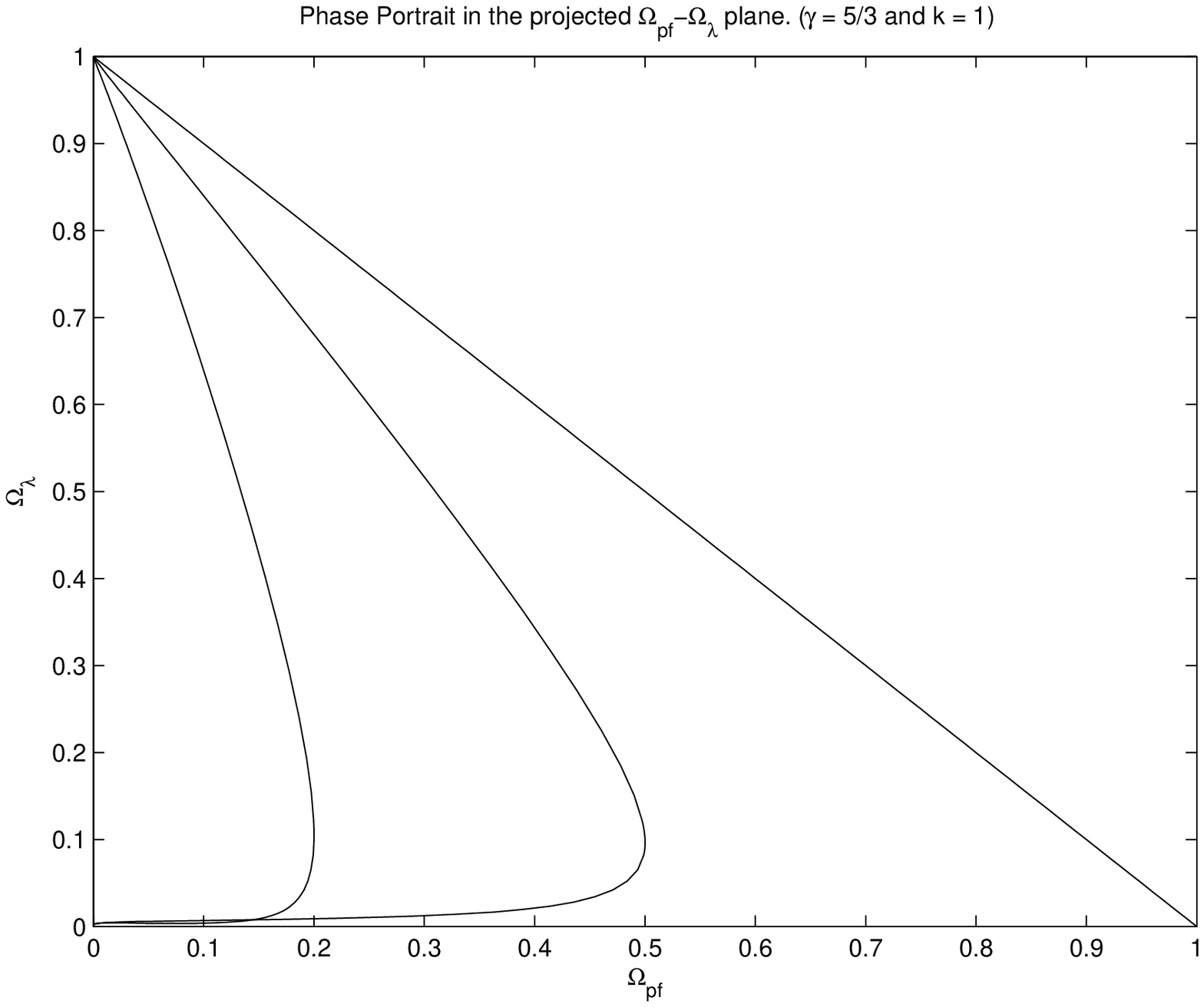}\qquad\includegraphics[width=7cm]{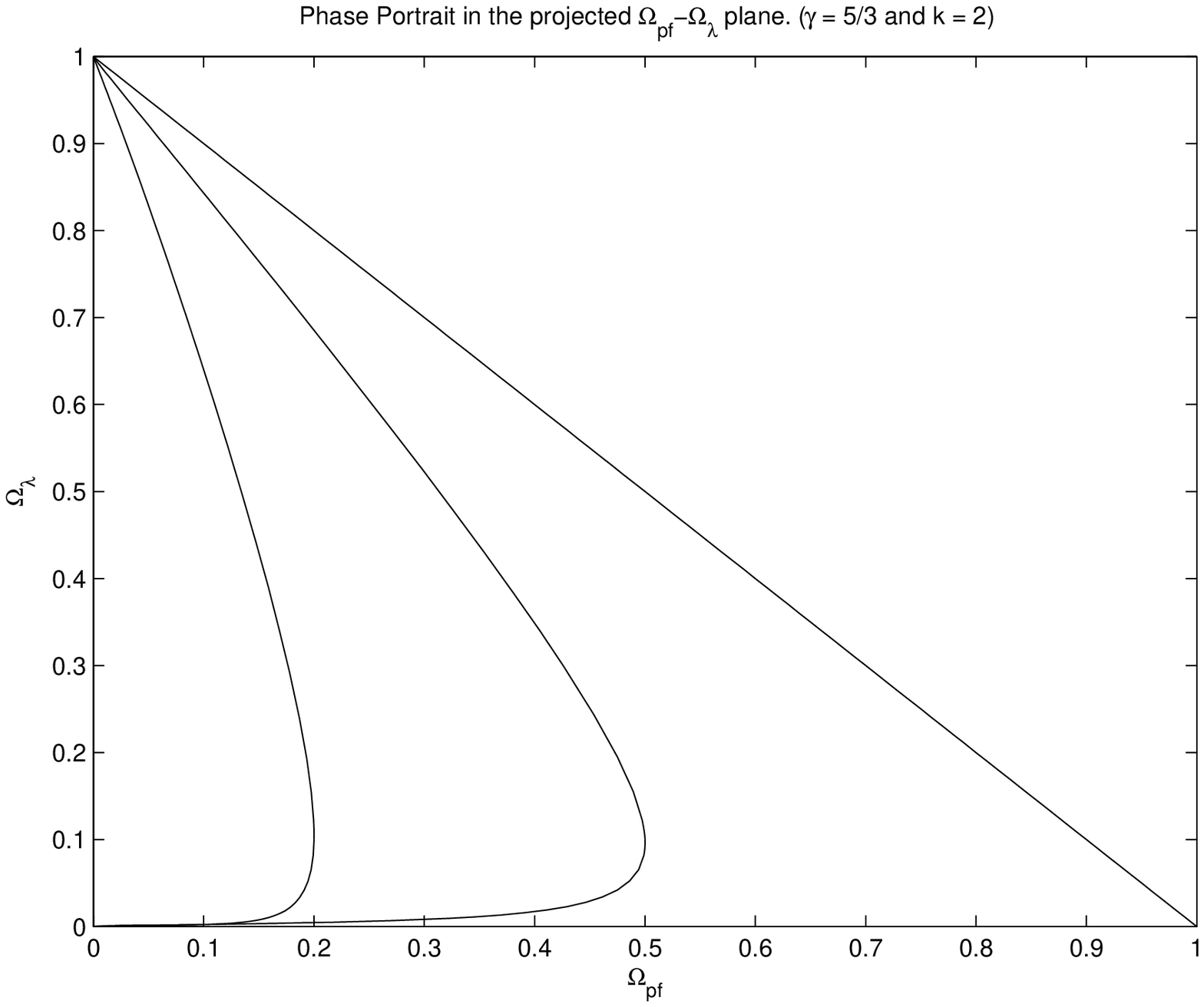}

\includegraphics[width=7cm]{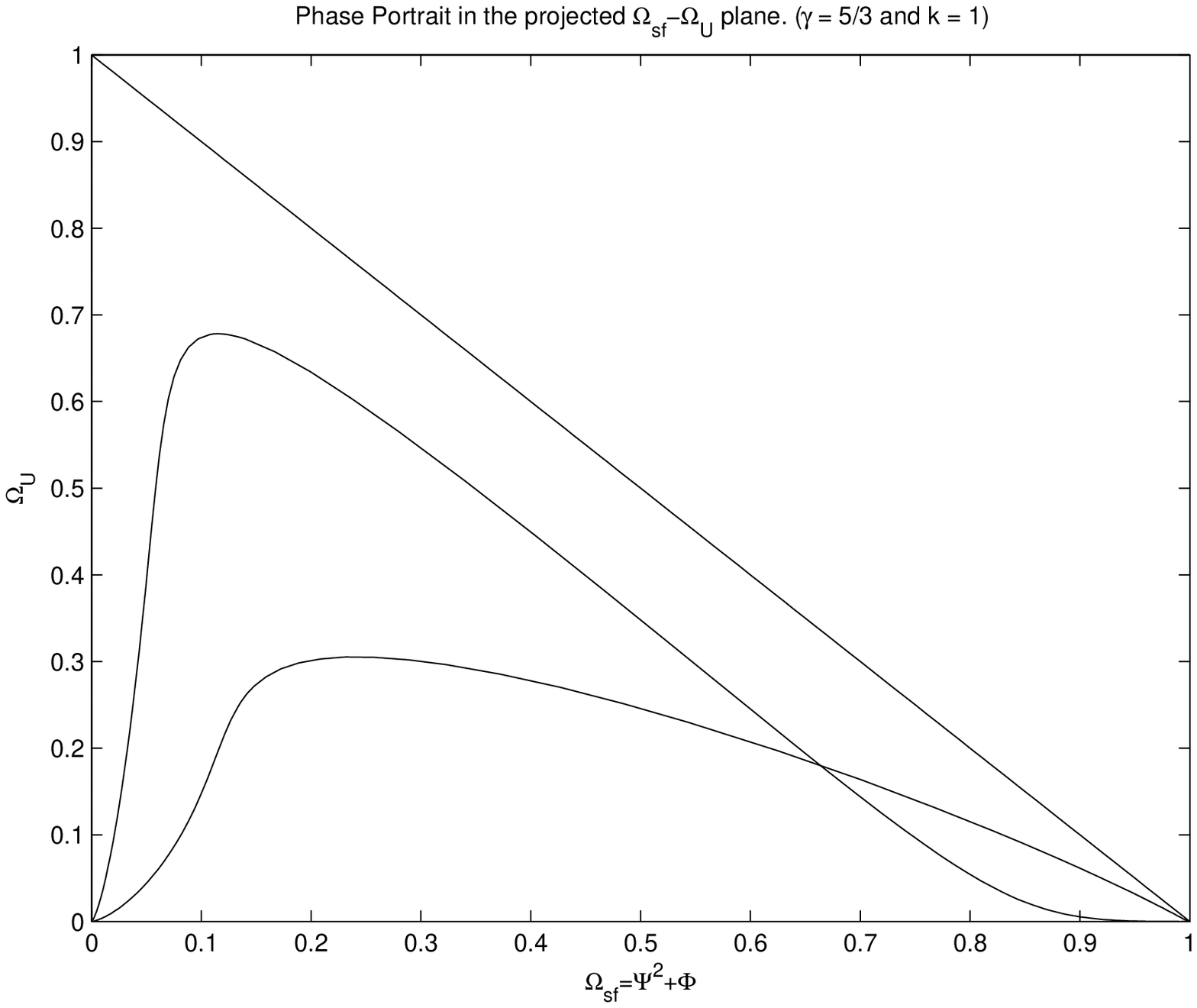}\qquad\includegraphics[width=7cm]{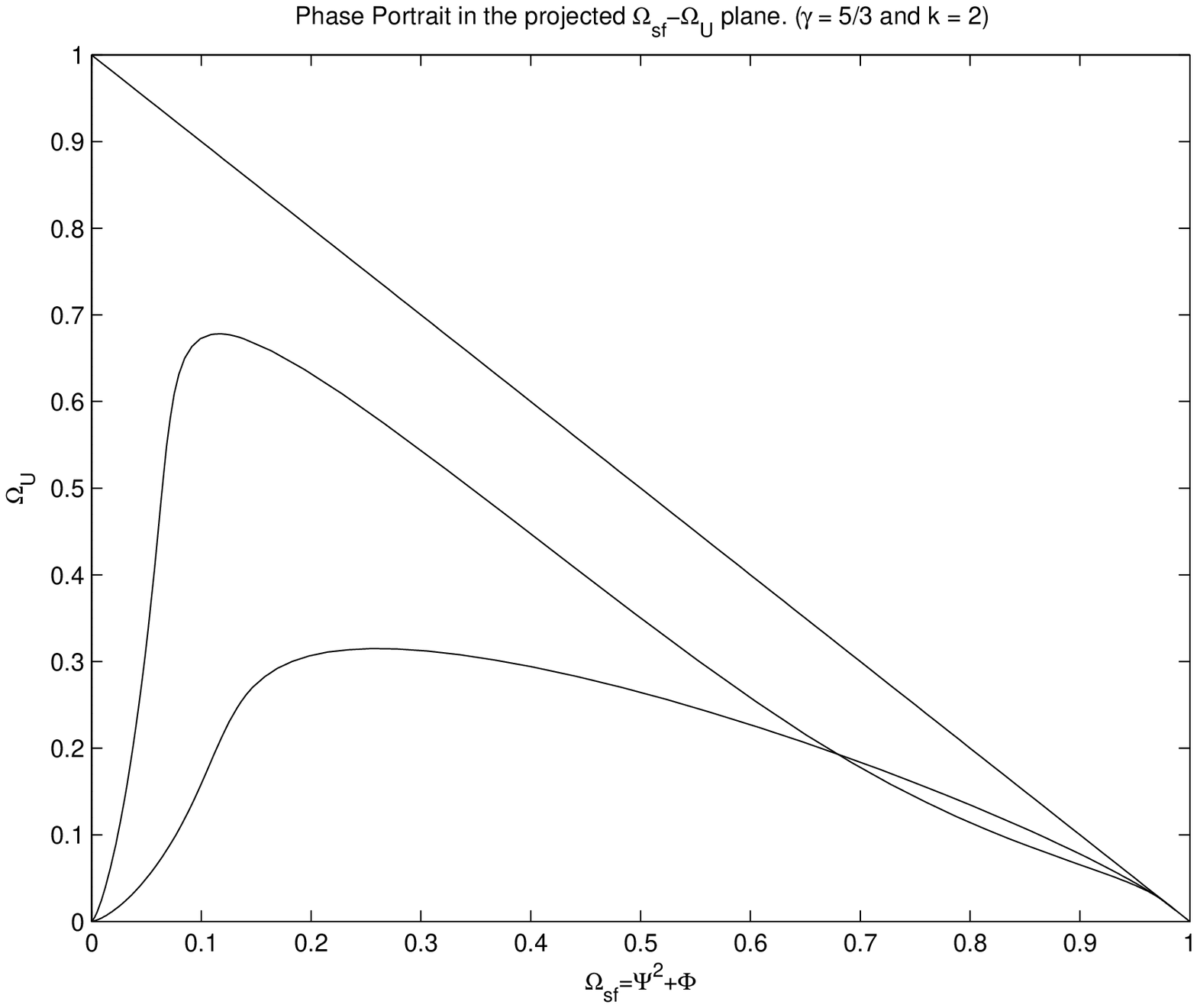}
\end{figure}

\begin{figure}[t]
\caption{Qualitative behaviour of trajectories when $\gamma=5/3>4/3$, $k=3$.}\label{figure6}
\includegraphics[width=7cm]{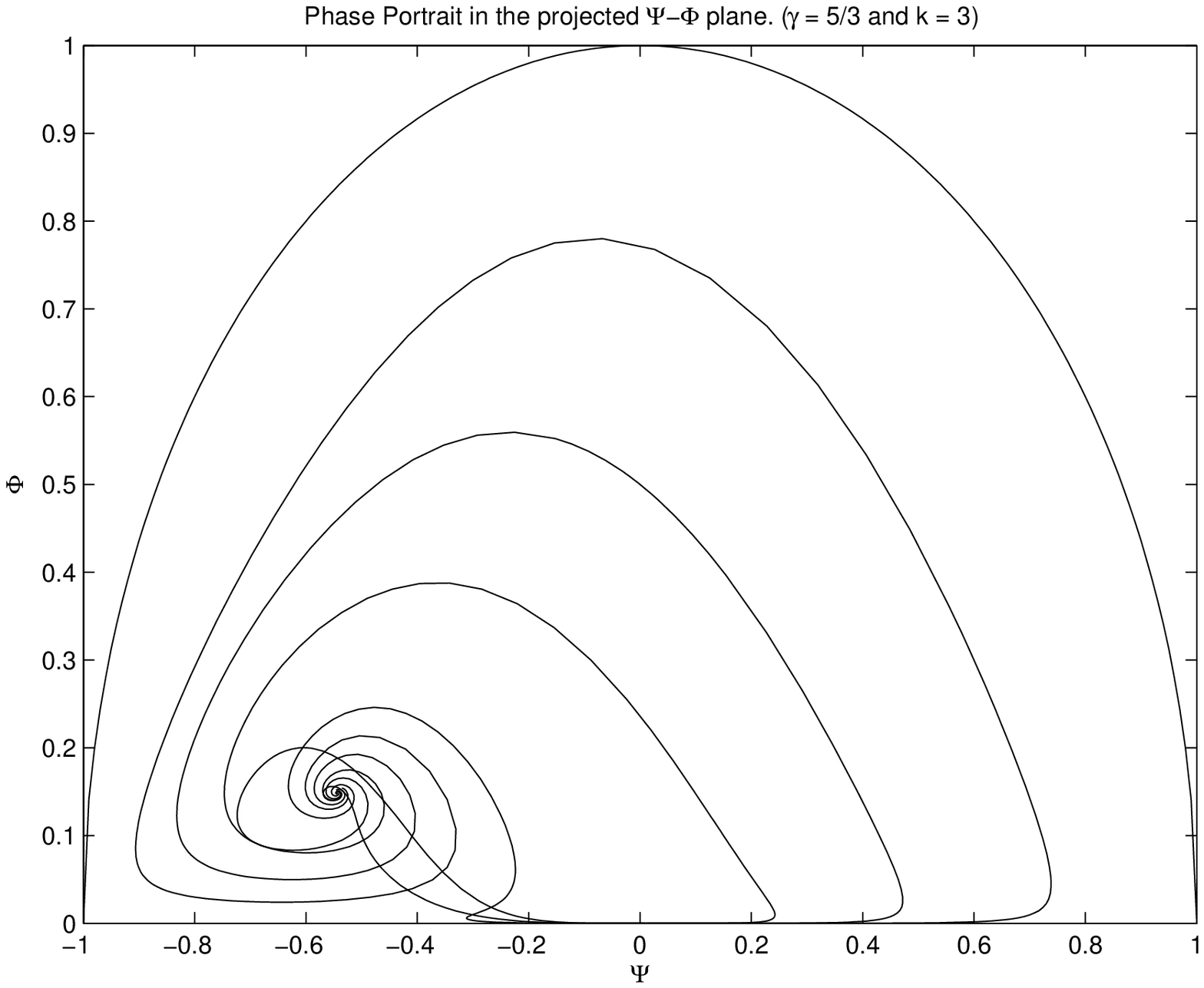} 

\includegraphics[width=7cm]{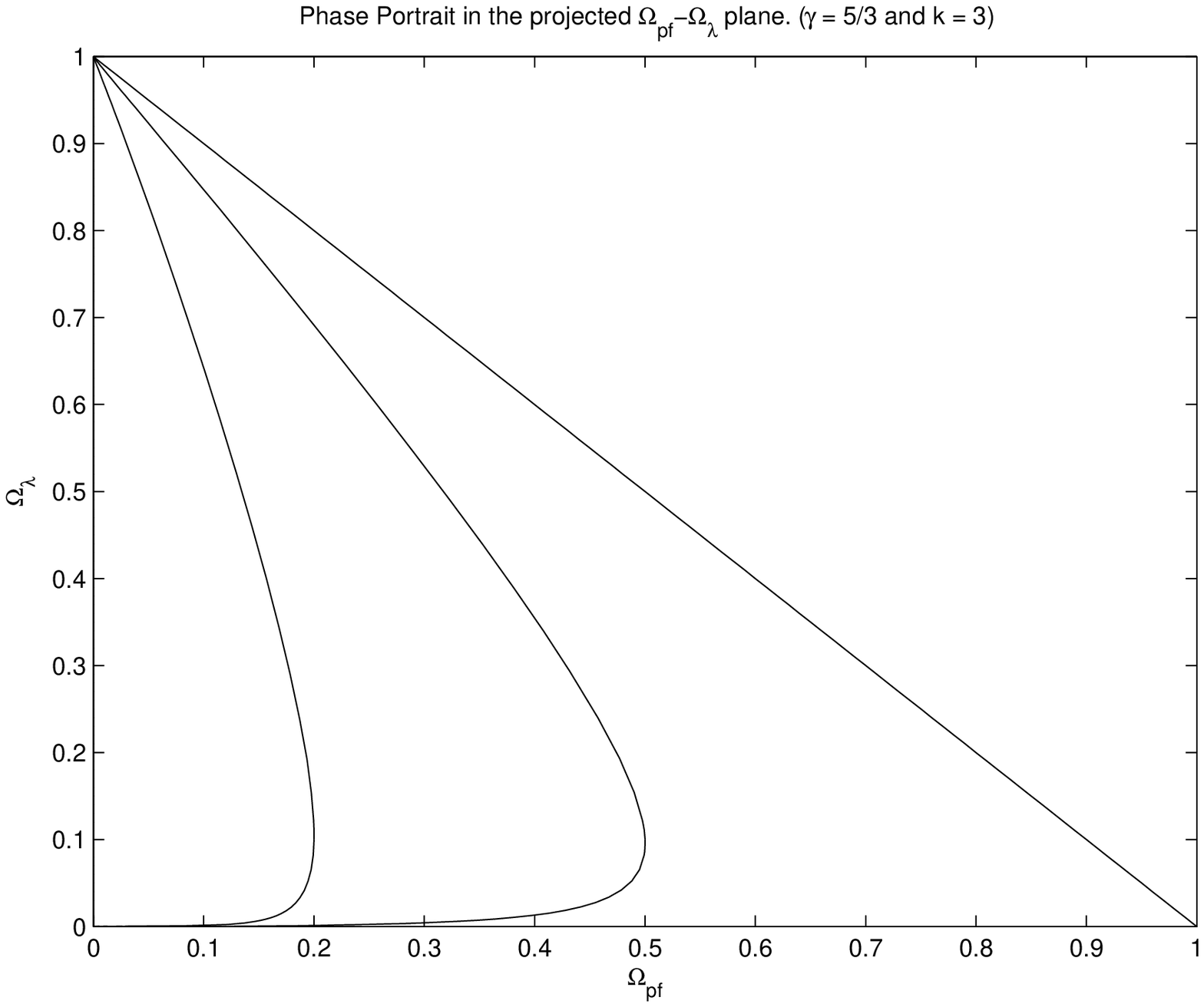} 

\includegraphics[width=7cm]{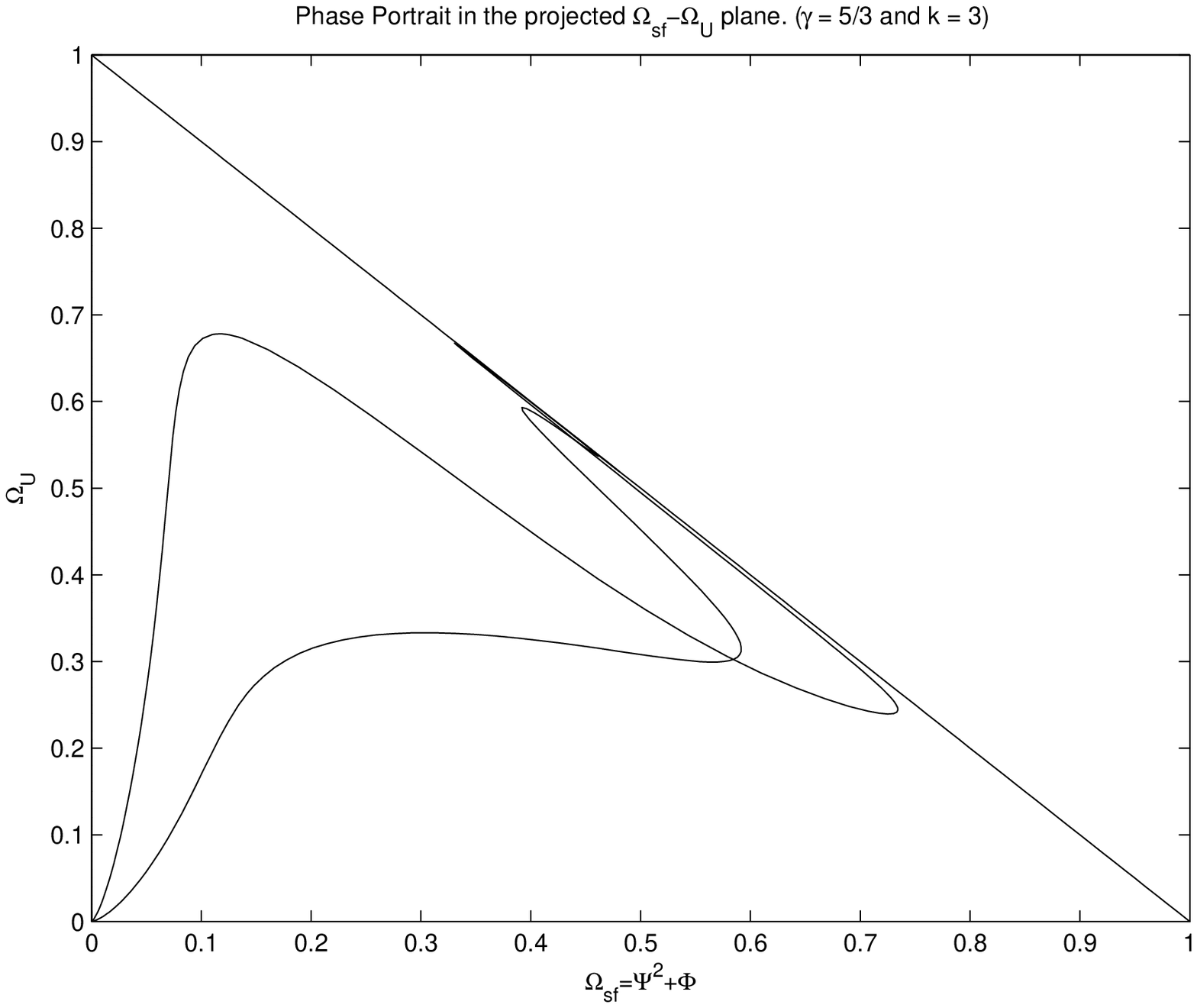} 
\end{figure}

\begin{description}
 
\item {$DS$;\ De-Sitter}\hfill\hfill \linebreak
The equilibrium point corresponding to the De Sitter solution in the original variables is represented by six different equilibrium points in the spherical coordinates defined above: $DS=[0,\tilde\theta,\tilde\phi,0,0,0,0,0]$ where $\tilde\theta=0,\pi/2,\pi$ and $\tilde \phi = 0,\pi/2$.   
The eigenvalues $ -3,-3,-1,-4,$ have an eigenspace spanned by $\Sigma_+,\Sigma_-,N,\Omega_{\cal U}$. The eigenvalues associated with the matter are $\lambda_{\tilde\rho}=-3\gamma/2$, $\lambda_{\Omega_{\lambda}}=-6\gamma$ when $\tilde\phi = 0$ and $\lambda_{\tilde\rho}=-3$, $\lambda_{\Omega_{\lambda}}=-12$ when $\tilde\phi = \pi/2$ except at the point $(\tilde\theta=\pi/2,\tilde\phi=\pi/2)$ where the eigenvalues are $\lambda_{\tilde\rho}=0$, $\lambda_{\Omega_{\lambda}}=0$. The eigenvalues associated with the two angular coordinates are: $\lambda_{\tilde\theta}=3$ when $(\tilde\theta=0,\pi,\tilde\phi=0,\pi/2)$ and   $\lambda_{\tilde\phi}=-3(2-\gamma)/2$ when $(\tilde\theta=0,\pi,\tilde\phi=0)$, and $\lambda_{\tilde\phi}=3(2-\gamma)/2$ when $(\tilde\theta=0,\pi,\tilde\phi=\pi/2)$;  $\lambda_{\tilde\theta}=-3$ when $(\tilde\theta=\pi/2,\tilde\phi=0,\pi/2)$ and   $\lambda_{\tilde\phi}=3\gamma/2$ when $(\tilde\theta=\pi/2,\tilde\phi=0)$, and $\lambda_{\tilde\phi}=-3\gamma/2$ when $(\tilde\theta=\pi/2,\tilde\phi=\pi/2)$.  The behaviour of the angular coordinates $\tilde\phi,\tilde\theta$ is illustrated in Figure \ref{figure1}.  We observe that all the eigenvalues not associated with the angular coordinates are negative, therefore, this point is a local attractor.  The dynamics in the $\tilde\phi-\tilde\theta$ plane indicate that $\tilde\phi\to\pi/2$ and $\tilde\theta\to\pi/2$ as $\tau \to \infty$.  This implies that this equilibrium point is strongly attracting along this direction, that is, this equilibrium point strongly attracts along the $\Omega_{pf}=\Psi=0$ direction in the original variables.  This local analysis, together with the global analysis in section \ref{monotonic}, implies that the De Sitter solution is a global attractor for a wide class of Bianchi II, Bianchi I, and zero curvature Robertson Walker cosmological models with a cosmological constant.

\item{$m$;\ Isotropic Braneworld Solution}\hfill\hfill \linebreak  
The equilibrium point $m$ in the original variables is represented by six different equilibrium points in the new variables,  $m=[0,\tilde\theta,\tilde\phi,0,0,0,1,0]$ where $\tilde\theta=0,\pi/2,\pi$ and $\tilde \phi = 0,\pi/2$.
For $\tilde\theta=0,\pi/2,\pi$ and $\tilde\phi=0$ the eigenvalues associated with the non-angular coordinates are $ \{6\gamma\},3\gamma/2,3(\gamma-1),3(\gamma-1),3\gamma-1,2(3\gamma-2)$.   For $\tilde\theta=0,\pi$ and $\tilde\phi=\pi/2$ the eigenvalues associated with the non-angular coordinates are $ \{12\},3 ,3,3,5,8$.  For $\tilde\theta=\pi/2$ and $\tilde\phi=\pi/2$ the eigenvalues associated with the non-angular coordinates are $ -3,-3,-1,0,\{0\},-4$.  The eigenvalues of the angular coordinates are the same as in De-Sitter case above (see Figure \ref{figure1}). We observe that this point is always a source for $\gamma>1$ and strongly repel away from the $\Omega_{pf}=\Phi=0$ direction.  As models evolve backwards in time, it is the kinetic energy of the scalar field that is the dominant component of the total energy density.  For $\gamma =1$, numerical analysis reveals that this point remains a source (see Figure \ref{figure2}).

\item{$R^0$;\ Robertson Walker Dark Radiation Solution}\hfill\hfill \linebreak  
The equilibrium point $R^0$ in the original variables is represented by six different equilibrium points in the new variables,  $R^0=[0,\tilde\theta,\tilde\phi,0,0,0,0,1]$ where $\tilde\theta=0,\pi/2,\pi$ and $\tilde \phi = 0,\pi/2$.
The eigenvalues $ -1,-1,1,\{4\}$ have an eigenspace spanned by $\Sigma_+,\Sigma_-,N,\Omega_{\cal U}$. The eigenvalues associated with the matter are $\lambda_{\tilde\rho}=(4-3\gamma)/2$, $\lambda_{\Omega_{\lambda}}=4-6\gamma$ when $\tilde\phi = 0$ and $\lambda_{\tilde\rho}=-1$, $\lambda_{\Omega_{\lambda}}=-8$ when $\tilde\phi = \pi/2$ except at the point $(\tilde\theta=\pi/2,\tilde\phi=\pi/2)$ where the eigenvalues are $\lambda_{\tilde\rho}=2$, $\lambda_{\Omega_{\lambda}}=4$.  The eigenvalues of the angular coordinates are the same as in De-Sitter case above.
We observe that the eigenvalues not associated with the angular coordinates are both positive and negative, therefore, this point is a saddle point.   If $\gamma<4/3$ then the point has a 3 dimensional unstable manifold in the rectangular coordinate system. (Note: If $\gamma<4/3$, then $\lambda_{\tilde\rho}=(4-3\gamma)/2>0$ along the invariant direction $\tilde\phi = 0$ but $\lambda_{\tilde\rho}=-1<0$ along the invariant directions $\tilde\theta=0,\tilde\phi = \pi/2$ and $\tilde\theta=0,\tilde\phi = \pi/2$.)   If $\gamma>4/3$, then the point has a 2 dimensional unstable manifold.

\item{$R^{II}$;\ Bianchi II Dark Radiation Solution}  \hfill\hfill \linebreak
The equilibrium point $R^{II}$ in the original variables is represented by six different equilibrium points in the new variables, $R^{II} =[0,\tilde\theta,\tilde\phi,\frac{1}{4},0,\frac{1}{4},0,\frac{7}{8}]$.
The eigenvalues $ -1,-1/2(1\pm\sqrt{6}i),\{4\}$  have an eigenspace spanned by $\Sigma_+,\Sigma_-,N,\Omega_{\cal U}$. The eigenvalues associated with the matter are $\lambda_{\tilde\rho}=(4-3\gamma)/2$, $\lambda_{\Omega_{\lambda}}=4-6\gamma$ when $\tilde\phi = 0$ and $\lambda_{\tilde\rho}=-1$, $\lambda_{\Omega_{\lambda}}=-8$ when $\tilde\phi = \pi/2$ except at the point $(\tilde\theta=\pi/2,\tilde\phi=\pi/2)$ where the eigenvalues are $\lambda_{\tilde\rho}=2$, $\lambda_{\Omega_{\lambda}}=4$.  The eigenvalues of the angular coordinates are the same as in De-Sitter case above.   If $\gamma<4/3$ then the equilibrium point has a 2 dimensional unstable manifold in the rectangular coordinate system.  If $\gamma>4/3$, then the equilibrium point has a one dimensional unstable manifold.  We also observe that if $\gamma>4/3$ then this point is a stable attractor in the $\Omega_{\Lambda}=0$ invariant set. 
 
\end{description}

\end{document}